\documentclass[12pt]{article}
\usepackage{amssymb, amsmath, hyperref, verbatim}

\def\ltwid{\mathrel{\raise.3ex\hbox{$<$\kern-.75em\lower1ex\hbox{$\sim$}}}}

\def\square{\kern1pt\vbox{\hrule height 1.2pt\hbox{\vrule width 1.2pt\hskip 3pt
   \vbox{\vskip 6pt}\hskip 3pt\vrule width 0.6pt}\hrule height 0.6pt}\kern1pt}

\begin{document}

\begin{titlepage}
\begin{flushright}
SPIN-08-38, ITP-UU-08/48, UFIFT-QG-08-06
\end{flushright}

\vspace{0.5cm}

\begin{center}
\bf{Infrared Propagator Corrections for Constant Deceleration}
\end{center}

\vspace{0.3cm}

\begin{center}
T. M. Janssen$^{*}$, S. P. Miao$^{**}$, T. Prokopec$^{\ddagger}$
\end{center}
\begin{center}
\it{Institute for Theoretical Physics \& Spinoza Institute,
Utrecht University,\\
Leuvenlaan 4, Postbus 80.195, 3508 TD Utrecht, THE NETHERLANDS}
\end{center}

\vspace{0.2cm}

\begin{center}
R. P. Woodard$^{\dagger\dagger}$
\end{center}
\begin{center}
\it{Department of Physics, University of Florida \\
Gainesville, FL 32611, UNITED STATES.}
\end{center}

\vspace{0.3cm}

\begin{center}
ABSTRACT
\end{center}
\hspace{0.3cm} We derive the propagator for a massless, minimally
coupled scalar on a $D$-dimensional, spatially flat, homogeneous and
isotropic background with arbitrary constant deceleration parameter.
Our construction uses the operator formalism, by integrating the
Fourier mode sum. We give special attention to infrared corrections
from the nonzero lower limit associated with working on finite
spatial sections. These corrections eliminate infrared divergences
that would otherwise be incorrectly treated by dimensional
regularization, resulting in off-coincidence divergences for those
special values of the deceleration parameter at which the infrared
divergence is logarithmic. As an application we compute the
expectation value of the scalar stress-energy tensor.

\vspace{0.3cm}

\begin{flushleft}
\end{flushleft}

\vspace{0.1cm}

\begin{flushleft}
$^{*}$ T.M.Janssen@uu.nl, \hspace{1cm} $^{**}$ S.Miao@uu.nl \\
$^{\ddagger}$ T.Prokopec@uu.nl, \hspace{1.2cm} $^{\dagger\dagger}$
woodard@phys.ufl.edu
\end{flushleft}

\end{titlepage}

\section{Introduction}

On the largest scales, the visible universe is well described by a
homogeneous, isotropic and spatially flat geometry whose invariant
element takes the form,
\begin{equation}
ds^2 \equiv g_{\mu\nu} dx^{\mu} dx^{\nu} = -dt^2 + a^2(t) d\vec{x}
\cdot d\vec{x} \; . \label{geom}
\end{equation}
Two derivatives of the scale factor $a(t)$ have great importance,
the Hubble parameter $H(t)$ and the deceleration parameter $q(t)$,
\begin{equation}
H(t) \equiv \frac{\dot{a}}{a} \qquad , \qquad q(t) \equiv -1 -
\frac{\dot{H}}{H^2} \equiv -1 + \epsilon(t) \; . \label{Hq}
\end{equation}
Although the Hubble parameter may have changed by as much as 57
orders of magnitude from primordial inflation to now, the
deceleration parameter is only thought to have varied from nearly
$-1$, during primordial inflation, to nearly $+1$, during the
epoch of radiation domination. And $q(t)$ has been approximately
constant for vast periods of cosmological evolution. It is
therefore interesting to consider physics during an epoch of
constant $q(t)$.

A field of great importance for primordial inflation is the
massless, minimally coupled scalar,
\begin{equation}
\mathcal{L} = -\frac12 \partial_{\mu} \varphi \partial_{\nu} \varphi
g^{\mu\nu} \sqrt{-g} = \frac12 \dot{\varphi}^2 a^{D-1} - \frac12
\Vert \vec{\nabla} \varphi \Vert^2 a^{D-3} \; . \label{Lag}
\end{equation}
This is true both as an approximation to the inflaton, on account of
the very flat potential, and also because dynamical gravitons have
the same kinetic operator \cite{Grish}. Note that we are assuming
the spatial manifold is $(D\!-\!1)$-dimensional in order to
facilitate the use of dimensional regularization.

The scalar propagator obeys the equation,
\begin{equation}
\sqrt{-g} \, \square i\Delta(x;x') \equiv \partial_{\mu} \Bigl(
\sqrt{-g} g^{\mu\nu} \partial_{\nu} i\Delta(x;x') \Bigr) = i
\delta^D(x \!-\! x') \; . \label{propeqn}
\end{equation}
An elegant solution to this equation has recently been obtained for
the case of arbitrary constant $\epsilon \equiv -\dot{H}/H^2$, which
of course means constant deceleration \cite{JP1}. This solution has
been used to compute the one loop correction to the effective field
equations in Einstein + Scalar \cite{JMP,JP2}. Unfortunately, the
solution diverges at certain values of $\epsilon$, even off
coincidence and with the dimensional regularization in effect. The
purpose of this paper is to elucidate the physical basis of this
divergence and to provide a very simple correction for it.

It is important to recognize that the propagator equation
(\ref{propeqn}) only fixes $i\Delta(x;x')$ up to the addition of a
homogeneous solution. There is no guarantee that a particular
solution can be interpreted as the expectation value of the
time-ordered product of $\varphi(x) \times \varphi(x')$ in the
presence of any state. One can see this even from the simple
harmonic oscillator of nonrelativistic quantum mechanics \cite{TW1}.
The Heisenberg position operator can be expressed in terms of the
initial position and momentum,
\begin{equation}
q(t) = q_0 \cos(\omega t) + \frac{p_0}{m \omega} \, \sin(\omega t)
\; .
\end{equation}
The associated propagator in the presence of an arbitrary state
$\vert \psi\rangle$ can be expressed in terms of three real numbers,
\begin{eqnarray}
\lefteqn{\langle \psi \vert T[q(t) q(t')] \vert \psi \rangle =
-\frac{i}{2 m \omega} \, \sin(\omega \vert t \!-\! t'\vert) }
\nonumber \\
& & \hspace{2cm} + \alpha \cos(\omega t) \cos(\omega t') + \beta
\sin[\omega (t \!+\! t')] + \gamma \sin(\omega t) \sin(\omega t') \;
. \qquad \label{HOprop}
\end{eqnarray}
The parameters $\alpha$, $\beta$ and $\gamma$ are,
\begin{equation}
\alpha \equiv \langle \psi \vert q_0^2 \vert \psi \rangle \quad ,
\quad \beta \equiv \frac1{2 m \omega} \ \langle \psi \vert q_0 p_0
\!+\! p_0 q_0 \vert \psi \rangle \quad , \quad \gamma \equiv
\frac1{m^2 \omega^2} \, \langle \psi \vert p_0^2 \vert \psi \rangle
\; .
\end{equation}
These parameters are not arbitrary. For example, the Uncertainty
Principle implies,
\begin{equation}
\alpha \gamma \geq \frac1{4 m^2 \omega^2} \; .
\end{equation}
However, the right hand side of (\ref{HOprop}) solves the propagator
equation for {\it any} choice of $\alpha$, $\beta$ and $\gamma$,
\begin{equation}
-m \Bigl(\frac{d^2}{dt^2} + \omega^2\Bigr) i\Delta(t;t') = i
\delta(t \!-\! t') \; .
\end{equation}

We seek a true propagator, rather than just a Green's function
that solves (\ref{propeqn}). To ensure this we construct
$i\Delta(x;x')$ from its operator mode sum in Section 2, assuming
that the spatial manifold is $\mathbb{R}^{D-1}$. This gives
precisely the result obtained previously \cite{JP1}. In Section 3
we note that the infinite space mode sum has an infrared
divergence for any value of $\epsilon \leq 2(D\!-\!1)/D$. In most
cases this is a power-law divergence that the automatic
subtraction of dimensional regularization simply but erroneously
discards. For certain special values of $\epsilon$ the divergence
becomes logarithmic, in which case the solution is infinite, even
off coincidence and for general $D$. In Section 4 we solve the
problem by formulating the mode sum on a finite spatial manifold.
This changes the propagator by the addition of a series of
homogeneous solutions which cancels the divergences. Section 5
uses the corrected propagator to compute the expectation value of
the scalar stress tensor. Our discussion comprises Section 6.

\section{Infinite Space Mode Sum}

The purpose of this section is to construct $i\Delta(x;x')$ from
the canonical operator formalism, assuming the spatial manifold is
$\mathbb{R}^{D-1}$. We begin by working out the Hubble parameter
and scale factor as functions of co-moving and conformal time.
Then we give the free field expansion of $\varphi(x)$. Taking the
expectation value of the time-ordered product of $\varphi(x)
\times \varphi(x')$ gives a Fourier integral expression for the
propagator which we are able to evaluate by analytically
continuing some integral identities. The section closes by noting
the special values of $\epsilon$ for which the resulting
propagator diverges, away from coincidence and with dimensional
regularization in effect.

It is simple to reconstruct the Hubble parameter and the scale
factor when $\epsilon$ is constant. If we define $H_0 \equiv H(0)$
and $a(0) \equiv 1$ then relation (\ref{Hq}) implies,
\begin{equation}
H(t) = \frac{H_0}{1 + \epsilon H_0 t} \qquad {\rm and} \qquad a(t) =
\Bigl[1 \!+\! \epsilon H_0 t\Bigr]^{\frac1{\epsilon}} \; .
\end{equation}
The conformal time $\eta$ is defined by $d\eta = dt/a(t)$. We can
choose the zero of conformal time so that the following relations
apply,
\begin{equation}
H(\eta) = \frac{H_0}{[-(1\!-\!\epsilon) H_0
\eta]^{\frac{-\epsilon}{1-\epsilon}}} \qquad {\rm and} \qquad
a(\eta) = \frac1{[-(1\!-\!\epsilon) H_0 \eta]^{\frac1{1
-\epsilon}}} \; ,
\end{equation}
For $0 \leq \epsilon < 1$ the universe is accelerating and
conformal time approaches zero from below; for $\epsilon > 1$ the
universe is decelerating and $\eta$ is positive with our
convention. A very useful relation is,
\begin{equation}
(1 \!-\! \epsilon) H a = -\frac1{\eta} \; .
\end{equation}

The massless, minimally coupled scalar on $\mathbb{R}^{D-1}$ can
be expressed as a Fourier integral of plane waves,
\begin{equation}
\varphi_{\infty}(\eta,\vec{x}) = \int \!
\frac{d^{D-1}k}{(2\pi)^{D-1}} \, \Bigl\{ u(\eta,k) e^{i\vec{k}
\cdot \vec{x}} \alpha(\vec{k}) + u^*(\eta,k) e^{-i\vec{k} \cdot
\vec{x}} \alpha^{\dagger}(\vec{k}) \Bigr\} \; . \label{free}
\end{equation}
Since $\varphi_\infty$ is a minimally coupled massless scalar, it
obeys the Klein Gordon equation $\square\varphi_\infty=0$. The
mode functions are quite complicated for general, time-dependent
$\epsilon(t)$ \cite{TW2}. However, for constant $\epsilon$ they
obey the simple equation
\begin{equation}
    \Big(-\partial_\eta^2-k^2+\frac{\nu^2-\frac{1}{4}}{\eta^2}\Big)\Big(a^{\frac{D}{2}-1}
    u(\eta,k)\Big)=0 ,\qquad \nu \equiv
\frac{D\!-\!1\!-\!\epsilon}{2(1\!-\!\epsilon)} \; ,
\end{equation}
where $k=\|\vec{k}\|$. The solution to this equation takes in this
case a simple form
\begin{equation}
u(\eta,k) = \sqrt{\frac{\pi |\eta|}{4}} \, a^{1-\frac{D}2}
H^{(1)}_{\nu}(k|\eta|).\label{udef}
\end{equation}
The creation and annihilation operators are canonically normalized,
\begin{equation} \Bigl[ \alpha(\vec{k}) , \alpha^{\dagger}(\vec{k}')
\Bigr] = (2\pi)^{D-1} \delta^{D-1}(\vec{k} \!-\! \vec{k}') \; .
\end{equation}
The state $\vert \Omega \rangle$ which is annihilated by
$\alpha(\vec{k})$ is known as {\it Bunch-Davies vacuum} \cite{BD}.
It corresponds to the Heisenberg state of minimum excitation in the
distant past.

The infinite space propagator is easy to write as a mode sum,
\begin{eqnarray}
\lefteqn{i\Delta_{\infty}(x;x') \equiv \Bigl\langle \Omega
\Bigl\vert \theta(t \!-\! t') \varphi_{\infty}(x)
\varphi_{\infty}(x') + \theta(t'\!-\!t) \varphi_{\infty}(x')
\varphi_{\infty}(x) \Bigr\vert
\Omega \Bigr\rangle \; , } \\
& & \hspace{-.4cm} = \frac{\pi}4 \, \sqrt{\eta \eta'} \, (a a')^{1
-\frac{D}2} \! \int \! \frac{d^{D-1}k}{(2\pi)^{D-1}} \,  e^{i
\vec{k}
\cdot (\vec{x} - \vec{x}')} \nonumber \\
& & \hspace{.1cm} \times \Biggl\{\theta(\Delta\eta)
H^{(1)}_{\nu}(k|\eta|) H^{(1)}_{\nu}(k|\eta'|)^* \!+
\theta(-\Delta\eta) H^{(1)}_{\nu}(k|\eta|)^*
H^{(1)}_{\nu}(k|\eta'|) \Biggr\} , \qquad
\end{eqnarray}
where $\Delta\eta=\eta-\eta'$.  Now recall the $D=4$ angular
integral,
\begin{equation}
\int \! \frac{d^3k}{(2\pi)^3} \, e^{i \vec{k} \cdot \Delta
\vec{x}} f(\Vert \vec{k}\Vert) = \frac{1}{2\pi^2} \int_0^{\infty}
\!\! dk \, k^2 \, \frac{\sin(k \Delta x)}{k \Delta x} \,f(k) \; .
\end{equation}
Here and henceforth we define $\Delta x \equiv \Vert \vec{x} \!-\!
\vec{x}'\Vert$. Generalizing to $D$ spacetime dimensions we have
$d^{D-1}k=k^{D-2}dk \,d\Omega_{D-2}$, $k=\Vert \vec{k}\Vert$, where
\begin{equation}
  d\Omega_{D-2}=\sin^{D-3}(\theta_{D-3})d\theta_{D-3}\sin^{D-4}(\theta_{D-4})d\theta_{D-4}\ldots
  d\phi,
\end{equation}
where $\theta_{D-3}$, $\theta_{D-4}$, $\ldots$, and $\phi$ are the
angles on the sphere $\mathbb{S}^{D-2}$. Making use of
\begin{equation}
\int d\Omega_{D-2}
          = \frac{2\pi^\frac{D-1}{2}}{\Gamma\left(\frac{D-1}{2}\right)}
          = \frac{2(4\pi)^{\frac{D}{2}-1}\Gamma\left(\frac{D}{2}\right)}
                 {\Gamma\left(D\!-\!1\right)} \,
\end{equation}
and Eq. (8.411.7) in~\cite{GR} gives,
\begin{equation}
\int \! \frac{d^{D-1}k}{(2\pi)^{D-1}} \, e^{i \vec{k} \cdot \Delta
\vec{x}} f(\Vert \vec{k}\Vert) = \frac1{2^{D-2} \pi^{\frac{D-1}2}}
\int_0^{\infty} \!\! dk \, k^{D-2} \, \frac{J_{\frac{D-3}2}(k
\Delta x)}{(\frac12 k \Delta x)^{\frac{D-3}2}} \, f(k) \; .
\end{equation}
Hence the scalar propagator can be reduced to a single integral,
\begin{eqnarray}
i\Delta_{\infty}(x;x') &=& \frac{\sqrt{\eta \eta'} \, (a
a')^{1 -\frac{D}2} }{ 2^D \pi^{\frac{D-3}2}} \! \int_{0}^{\infty}
\! dk \, k^{D-2} \, \frac{J_{\frac{D-3}2}(k \Delta x)}{(\frac12 k
\Delta x)^{\frac{D-3}2}}
\nonumber \\
& \times&\! \biggl\{\theta(\Delta \eta) H^{(1)}_{\nu}(k|\eta|)
H^{(1)}_{\nu}(k|\eta'|)^* \!+ \theta(-\Delta \eta)
H^{(1)}_{\nu}(k|\eta|)^* H^{(1)}_{\nu}(k|\eta'|) \biggr\} . \qquad
\end{eqnarray}
To reach the final form we make the change of variable,
\begin{equation}
z \equiv \sqrt{\eta \eta'} \, k \; \Longrightarrow \; k \Delta x =
\frac{\Delta x}{\sqrt{\eta \eta'}} \, z \quad , \quad k |\eta| =
\sqrt{\frac{\eta}{\eta'}} \, z \quad , \quad k |\eta'| =
\sqrt{\frac{\eta'}{\eta}} \, z \; .
\end{equation}
Recalling that $1/(\eta a) = -(1\!-\!\epsilon) H$ gives,
\begin{eqnarray}
\lefteqn{i\Delta_{\infty}(x;x') = \frac{(\eta \eta' \, a a')^{1
-\frac{D}2} }{ 2^D \pi^{\frac{D-3}2}} \! \int_{0}^{\infty} \! dz \,
z^{D-2} \times \frac{J_{\frac{D-3}2}\Bigl(\frac{\Delta x}{\sqrt{\eta
\eta'}} \, z\Bigr)}{ \Bigl(\frac12 \frac{\Delta x}{\sqrt{\eta
\eta'}} \, z\Bigr)^{\frac{D-3}2}} }
\nonumber \\
& & \hspace{.8cm} \times \Biggl\{\theta(\Delta \eta)
H^{(1)}_{\nu}\Bigl( \sqrt{\frac{\eta}{\eta'}} \, z \Bigr)
H^{(1)}_{\nu}\Bigl(\sqrt{\frac{\eta'}{ \eta}} \, z\Bigr)^* +
\theta(-\Delta \eta) \times \Bigl({\rm conjugate}\Bigr)
\Biggr\} , \qquad \\
& & \hspace{-.3cm} = \frac{\Bigl[ (1\!-\!\epsilon)^2 H
H'\Bigr]^{\frac{D}2-1}}{ (4 \pi)^{\frac{D}2}} \times
\frac{\pi^{\frac32} \, 2^{\frac{D-3}2}}{ \Bigl(\frac{\Delta
x}{\sqrt{\eta \eta'}}\Bigr)^{\frac{D-3}2}} \times \int_{0}^{\infty}
\! dz \, z^{\frac{D-1}2} \times
J_{\frac{D-3}2}\Bigl(\frac{\Delta x}{\sqrt{\eta \eta'}} \, z\Bigr) \nonumber \\
& & \hspace{.8cm} \times \Biggl\{\theta(\Delta \eta)
H^{(1)}_{\nu}\Bigl( \sqrt{\frac{\eta}{\eta'}} \, z \Bigr)
H^{(1)}_{\nu}\Bigl(\sqrt{\frac{\eta'}{ \eta}} \, z\Bigr)^* +
\theta(-\Delta \eta) \times \Bigl({\rm conjugate}\Bigr) \Biggr\} .
\qquad \label{modesum}
\end{eqnarray}

By analytic continuation of relation (6.578.10) and employing
relations (8.702), (8.407.1), (8.476.10) and (9.131.1) in
\cite{GR} one can show,
\begin{eqnarray}
\lefteqn{\int_0^{\infty} \! dx \, x^{\mu+1} J_{\mu}(c x)
H^{(1)}_{\nu}(a x) H^{(1)}_{\nu}(b x)^* = \frac{\Gamma(\mu \!+\! 1
\!+\! \nu) \Gamma(\mu \!+\!
1 \!-\! \nu)}{\pi^{\frac32} \, \Gamma(\mu \!+\! \frac32)} } \nonumber \\
& & \hspace{2.6cm} \times \frac{(\frac12 c)^{\mu}}{(a b)^{\mu+1}} \,
\mbox{}_2 F_1\Biggl(\mu \!+\! 1 \!+\! \nu, \mu \!+\!1 \!-\! \nu ;
\mu \!+\! \frac32 ; \frac{(a \!+\! b)^2 \!-\! c^2}{4 a b}\Biggr) \;
. \qquad \label{guess}
\end{eqnarray}
Now make the following assignments for the various parameters in
(\ref{guess}),
\begin{equation}
\mu \longrightarrow \frac{D\!-\!3}2 \quad , \quad a \longrightarrow
\sqrt{\frac{\eta}{\eta'}} \quad , \quad b \longrightarrow
\sqrt{\frac{\eta'}{\eta}} \quad {\rm and} \quad c \longrightarrow
\frac{\Delta x}{\sqrt{\eta \eta'}} \; .
\end{equation}
With these assignments we have,
\begin{equation}
\frac{(a \!+\! b)^2 \!-\! c^2}{4 a b} = 1 - \Bigl( \frac{c^2 \!-\!
(a \!-\! b)^2}{4 a b}\Bigr) \longrightarrow 1 - \Bigl(\frac{\Delta
x^2 - \Delta \eta^2}{4 \eta \eta'}\Bigr) \; .
\end{equation}
Hence the mode sum (\ref{modesum}) for the propagator can be given
the simple spacetime expression,
\begin{eqnarray}
i\Delta_{\infty}(x;x') = \frac{\Bigl[ (1\!-\!\epsilon)^2 H
H' \Bigr]^{\frac{D}2 -1} }{(4 \pi)^{\frac{D}2}}
\frac{\Gamma(\frac{D-1}2 \!+\!\nu)
\Gamma(\frac{D-1}2 \!-\! \nu)}{\Gamma(\frac{D}2)}
\; \mbox{}_2 F_1\biggl(\frac{D\!-\!1}2 \!+\! \nu,
\frac{D\!-\!1}2 \!-\! \nu ; \frac{D}2 ; 1 \!-\! \frac{y}4\biggr)
\, .\hskip -0.6cm
\nonumber\\
 \label{Tomislav}
\end{eqnarray}
Here and henceforth we define the quantity $y=y(x;x')$ as,
\begin{equation}
y(x;x') \equiv \frac{\bigl\Vert \vec{x} \!-\! \vec{x}\,'
\bigr\Vert^2 - \bigl( \vert \eta \!-\! \eta' \vert \!-\! i
\varepsilon \bigr)^2}{\eta\eta'} \; . \label{ydef}
\end{equation}
Note the distinction between the infinitesimal quantity
$\varepsilon$, used to define the pole prescription, and the
parameter $\epsilon \equiv -\dot{H}/H^2$. The propagator
(\ref{Tomislav}) is the generalization of the Chernikov-Tagirov
propagator for de Sitter space to space-times with constant, but
arbitrary $\epsilon$~\cite{Chernikov:1968zm}. The constant
$\epsilon$ propagator was already found for $D=4$ in
~\cite{Bunch:1977sq}.

Expression (\ref{Tomislav}) is precisely the result that was
obtained previously by solving the propagator equation
(\ref{propeqn}) with the ansatz of $(H H')^{\frac{D}2-1}$ times a
function of $y(x;x')$ \cite{JMP,JP2}. If we employ the
transformation formulae for hypergeometric functions and then
their series expansion we can write this as,
\begin{eqnarray}
i\Delta_{\infty}(x;x')
&=& \frac{\Bigl[ (1\!-\!\epsilon)^2 H H'
\Bigr]^{\frac{D}2 -1} }{(4 \pi)^{\frac{D}2}} \Biggl\{
\Gamma\Bigl(\frac{D}2\!-\!1\Bigr) \Bigl(
\frac{4}{y}\Bigr)^{\frac{D}2 -1} \mbox{}_2 F_1\Bigl(\frac12
\!+\!\nu,
\frac12 \!-\! \nu ; 2 \!-\! \frac{D}2 ; \frac{y}4\Bigr) \qquad \nonumber \\
& & + \frac{\Gamma(\frac{D-1}2 \!+\!\nu) \Gamma(\frac{D-1}2 \!-\!
\nu) \Gamma(1 \!-\! \frac{D}2)}{\Gamma(\frac12 \!+\! \nu)
\Gamma(\frac12 \!-\! \nu)} \, \mbox{}_2 F_1\Bigl(\frac{D-1}2 \!+\!
\nu , \frac{D-1}2 \!-\! \nu ;
\frac{D}2 ; \frac{y}4\Bigr) \Biggr\} , \qquad \\
& & \hspace{-.5cm} = \frac{\Bigl[ (1\!-\!\epsilon)^2 H H'
\Bigr]^{\frac{D}2 -1} }{(4 \pi)^{\frac{D}2}}
\Gamma\Bigl(\frac{D}2\!-\!1\Bigr) \Biggl\{ \Bigl(
\frac{4}{y}\Bigr)^{\frac{D}2 - 1} + \frac{\Gamma(2 \!-\!
\frac{D}2)}{
\Gamma(\frac12 \!+\! \nu) \Gamma(\frac12 \!-\! \nu)} \nonumber \\
& & \times \sum_{n=0}^{\infty} \Biggl[ \frac{\Gamma(\frac32 \!+\!
\nu \!+\! n) \Gamma(\frac32 \!-\! \nu \!+\! n)}{\Gamma(3 \!-\!
\frac{D}2 \!+\! n) \,
(n \!+\! 1)!} \Bigl(\frac{y}4\Bigr)^{n - \frac{D}2 +2} \nonumber \\
& & \hspace{5cm} - \frac{\Gamma(\frac{D-1}2 \!+\! \nu \!+\! n)
\Gamma(\frac{D-1}2 \!-\! \nu \!+\! n)}{\Gamma(\frac{D}2 \!+\! n) \,
n!} \Bigl(\frac{y}4\Bigr)^n \Biggr] \Biggr\} . \qquad
\label{problem}
\end{eqnarray}
The problem with (\ref{problem}) is that the gamma functions on
the last line diverge for certain values of $\epsilon$,
irrespective of whether or not $x^{\prime \mu} = x^{\mu}$ and with
the dimensional regularization still in effect. For the
inflationary case $0 \leq \epsilon < 1$ divergences occur at,
\begin{equation}
\epsilon = \frac{2N}{D \!-\! 2 \!+\! 2N} \Longrightarrow
\Gamma\Bigl( \frac{D\!-\!1}2 - \nu + n\Bigr) = \Gamma(-N + n)
\quad {\rm for} \quad N = 0,1,2,\dots
\label{lower}
\end{equation}
 For the decelerating case of $1 < \epsilon$ divergences are found
at,
\begin{equation}
\epsilon = 2\frac{D \!-\! 1 \!+\! N}{D \!+\! 2N} \Longrightarrow
\Gamma\Bigl(\frac{D \!-\!1}2 + \nu + n\Bigr) = \Gamma(-N + n)
\quad {\rm for} \quad N = 0,1,2,\dots
\label{upper}
\end{equation}

\section{Origin of the Problem}

The purpose of this section is to explain why the infinite space
propagator (\ref{problem}) diverges for the special values of
$\epsilon$ given in (\ref{lower}-\ref{upper}). We begin by
discussing infrared divergences of the mode sum (\ref{modesum}). The
discrete values of $\epsilon$ result from how these infrared
divergences are handled by dimensional regularization. Of course an
infrared divergence {\it should not} be subtracted like an
ultraviolet divergence! The correct procedure is to instead identify
and remove whatever unphysical feature led to the infrared
divergence. We close by doing this for the scalar propagator.

It has long been known that the infinite space mode sum
(\ref{modesum}) has infrared divergences for broad ranges of
constant $\epsilon$ \cite{FP}. They follow from the small argument
expansions of the Bessel function,
\begin{equation}
J_{\nu}(z) = \sum_{n=0}^{\infty} \frac{(-1)^n (\frac12 z)^{\nu +
2n}}{n! \Gamma(\nu \!+\! n \!+\! 1)} \; ,
\end{equation}
and from its relation to the Hankel function,
\begin{equation}
H^{(1)}_{\nu}(z) \equiv J_{\nu}(z) + i N_{\nu}(z) =
\frac{i}{\sin(\nu \pi)} \Bigl\{ e^{-i \nu \pi} J_{\nu}(z) -
J_{-\nu}(z)\Bigr\} \; . \label{Hankel}
\end{equation}
The leading small $z$ behavior for the terms on the first line of
(\ref{modesum}) is universal,
\begin{equation}
z^{\frac{D-1}2} \times J_{\frac{D-3}2}\Bigl(\frac{\Delta
x}{\sqrt{\eta \eta'}} \, z \Bigr) \longrightarrow
\frac{z^{D-2}}{\Gamma(\frac{D-1}2)} \times \Bigl(\frac{\Delta x}{2
\sqrt{\eta \eta'}} \Bigr)^{\frac{D-3}2} \; . \label{L1}
\end{equation}
What we get for the two Hankel functions on the second line depends
upon whether the universe is accelerating or decelerating. For the
inflationary case of $0 \leq \epsilon < 1$ the index $\nu$ is
positive and it is the $J_{-\nu}$ terms in the Hankel functions
which make the leading contributions for small $z$,
\begin{equation}
H^{(1)}_{\nu}\Bigl(\sqrt{\frac{\eta}{\eta'}} \, z \Bigr)
H^{(1)}_{\nu}\Bigl(\sqrt{\frac{\eta'}{\eta}} \, z\Bigr)^*
\longrightarrow \frac{(\frac12 z)^{-2\nu}}{\sin^2(\nu \pi)
\Gamma^2(-\nu \!+\! 1)} = \frac{2^{2\nu} \Gamma^2(\nu)}{\pi^2}
\times z^{-2\nu} \; . \label{L2}
\end{equation}
For the decelerating case of $1 < \epsilon \leq (D\!-\!1)$ the index
$\nu$ is negative and it is the $J_{+\nu}$ terms in the Hankel
functions which make the leading contributions for small $z$,
\begin{equation}
H^{(1)}_{\nu}\Bigl(\sqrt{\frac{\eta}{\eta'}} \, z \Bigr)
H^{(1)}_{\nu}\Bigl(\sqrt{\frac{\eta'}{\eta}} \, z\Bigr)^*
\longrightarrow \frac{(\frac12 z)^{2\nu}}{\sin^2(\nu \pi)
\Gamma^2(\nu \!+\! 1)} = \frac{2^{-2\nu} \Gamma^2(-\nu)}{\pi^2}
\times z^{2\nu} \; . \label{L3}
\end{equation}

From the preceding discussion we see that the asymptotic small $z$
form of the integrand in (\ref{modesum}) is a constant times,
\begin{equation}
z^{D-2 - 2\vert \nu\vert} \; .
\end{equation}
For this to produce an infrared divergence requires the exponent to
be $-1$ or less. For inflationary case of $0 \leq \epsilon < 1$ we
have $\nu > 0$ and the condition for a divergence is always met,
\begin{equation}
0 \leq \epsilon < 1 \quad \Longrightarrow \quad -1 \geq D \!-\!2 -
\frac{D\!-\!1 \!-\! \epsilon}{1 \!-\! \epsilon} = -1 -
\frac{\epsilon (D\!-\!2)}{1 \!-\! \epsilon} \; .
\end{equation}
For the decelerating case of $1 < \epsilon < (D\!-\!1)$ we have $\nu
< 0$ and the condition is met as long as $\epsilon \leq
2(D\!-\!1)/D$,
\begin{equation}
1 < \epsilon \leq \frac{2(D\!-\!1)}{D} \quad \Longrightarrow \quad
-1 \geq D \!-\!2 + \frac{D\!-\!1 \!-\! \epsilon}{1 \!-\! \epsilon}
= -1 - \frac{2 (D\!-\!1) \!-\! D\epsilon }{\epsilon \!-\! 1} \; .
\end{equation}

Although the mode sum (\ref{modesum}) has infrared divergences for
all $0 \leq \epsilon \leq 2(D\!-\!1)/D$, the final result
(\ref{problem}) only diverges for the discrete values of $\epsilon$
given by expressions (\ref{lower}-\ref{upper}). This is because
dimensional regularization \cite{HV,BG} automatically subtracts
power law divergences and only registers logarithmic divergences.
For most values of $\epsilon$ the infrared divergence is a power
law, and dimensional regularization --- quite incorrectly --- sets
it to zero. It is only for the discrete values
(\ref{lower}-\ref{upper}) that a logarithmic divergence occurs and
causes expression (\ref{problem}) to become ill-defined. To see
this, note that the logarithmic divergence could derive from any of
the order $z^{2N}$ series corrections to the leading small $z$ term.
For the inflationary range of $0 \leq \epsilon < 1$ the condition
for a logarithmic infrared divergence corresponds precisely to
(\ref{lower}),
\begin{eqnarray}
0 \leq \epsilon < 1 & \Longrightarrow & (D\!-\!2) + 2N -
\frac{D\!-\!1
\!-\! \epsilon}{1 \!-\! \epsilon} = -1 \; , \\
& \Longrightarrow & \epsilon = \frac{2 N}{D \!-\! 2 \!+\! 2 N} \quad
{\rm for} \quad N = 0,1,2,\dots
\label{poles:lowerN}
\end{eqnarray}
In $D=4$ dimensions the problem values are $\epsilon =
0,\frac12,\frac23, \frac34,\dots$ For the decelerating case of $1 <
\epsilon \leq 2(D\!-\!1)/D$ the condition for a logarithmic infrared
divergence agrees with (\ref{upper}),
\begin{eqnarray}
1 < \epsilon \leq \frac{2 (D\!-\!1)}{D} & \Longrightarrow &
(D\!-\!2) + 2N +
\frac{D\!-\!1 \!-\! \epsilon}{1 \!-\! \epsilon} = -1 \; , \\
& \Longrightarrow & \epsilon = 2 \,\frac{D \!-\! 1 \!+\! N}{D
\!+\! 2 N} \quad {\rm for} \quad N = 0,1,2,\dots
\label{poles:upperN}
\end{eqnarray}
For $D=4$ this corresponds to $\epsilon
=\frac32,\frac43,\frac54,\dots$

It is important to understand that the infinite space propagator has
physical problems for {\it every} value of $\epsilon$ in the
infrared divergent range $0 \leq \epsilon \leq 2(D\!-\!1)/D$,
whether or not $\epsilon$ happens to take one of the critical values
(\ref{lower}-\ref{upper}) necessary for a logarithmic divergence.
This is because ultraviolet and infrared divergences mean different
things. Ultraviolet divergences indicate that loop corrections have
made an infinite change between observed parameters and the
corresponding parameters of the Lagrangian. They can be canceled by
expressing the parameters of the Lagrangian in terms of observed
quantities plus counterterms which subtract off the divergences
\cite{SW}. The automatic subtraction of dimensional regularization
is not an error for ultraviolet divergences; it merely saves one the
trouble of defining and subtracting the appropriate counterterm to
cancel a power law divergence.

Infrared divergences do not mean anything about parameters in the
Lagrangian. Instead, they signify that there is something wildly
unphysical about the computation being done. One does not deal with
an infrared divergence by subtracting a counterterm; the correct
procedure is rather to compute physically well-defined quantities.
The classic example is the Bloch-Nordsieck switch from infrared
divergent, exclusive processes to infrared finite, inclusive
processes in quantum electrodynamics \cite{SW}. It will be seen that
employing the automatic subtraction of dimensional regularization to
remove a power law infrared divergence corresponds to adding an
illegal counterterm to make an unphysical question return a finite
answer, instead of reformulating the question in more physical
terms.

The unphysical thing about the infinite space propagator is that a
local observer cannot prepare the initially super-horizon modes of
the state in coherent Bunch-Davies vacuum. Two plausible fixes have
been proposed:
\begin{itemize}
\item{One could work on infinite space as in (\ref{free}) but assume
that the super-horizon modes are less singular than Bunch-Davies
vacuum \cite{AV}. Because only the super-horizon modes change
there would be no effect on the Hadamard behavior of the
propagator. If one continues to regard the state as obeying
$\alpha(\vec{k}) \vert \Omega \rangle = 0$, this fix corresponds
to changing the super-horizon mode functions $u(t,k)$ from the
Bunch-Davies choice (\ref{udef}). Of course their time dependence
is determined by the scalar field equation but their initial
values and those of their first time derivatives can be freely
specified. For example, if these initial values were chosen to be
those of the Bunch-Davies mode functions for $\epsilon = D/2$
(regardless of the actual value of $\epsilon$) then there would be
no infrared divergence, either initially or at any later
time\cite{Fulling:1978ht}.}
\item{One could also work on a compact spatial manifold such as
a torus $T^{D-1}$ for which there are no initially super-horizon modes
\cite{TW3}. In this case the free field expansion becomes a sum
rather than an integral but it is generally valid to make the
integral approximation to this sum, with a nonzero lower limit. When
this was done for the graviton propagator on de Sitter background
($\epsilon = 0$) there is no disturbance to powerful consistency
checks such as the one loop Ward identity \cite{TW4} and the nature
of allowed counterterms \cite{MW,KW}. The renormalization of scalar
field theories is not even affected at two loop order
\cite{OW,BOW,PTW}.}
\end{itemize}

\section{Finite Space Mode Sum}

The purpose of this section is to implement the second of the two
fixes described above: the one based upon a finite-sized spatial
manifold \cite{TW3}. We show how this changes the mode sum for the
propagator. We also derive the corrections it makes to the
integrated, position-space form. Explicit demonstrations are given
that the correction terms cure the $N=0$ and $N=1$ divergences in
expressions (\ref{lower}-\ref{upper}). And certain special cases are
checked against known results \cite{OW,ITTW}.

We work on $T^{D-1}$, which supports the spatially flat FRW
geometry (\ref{geom}). If the coordinate radius in each direction
is $2\pi/k_0$ then the integral approximation for the free field
expansion of the operator is the same as (\ref{free}) except that
the integral is cut off at $\Vert \vec{k} \Vert = k_0$,
\begin{equation}
\varphi(t,\vec{x}) = \int \! \frac{d^{D-1}k}{(2\pi)^{D-1}} \,
\theta(k - k_0) \Bigl\{ u(t,k) e^{i\vec{k} \cdot \vec{x}}
\alpha(\vec{k}) + u^*(t,k) e^{-i\vec{k} \cdot \vec{x}}
\alpha^{\dagger}(\vec{k}) \Bigr\} \; .
\end{equation}
Of course the same cutoff works its way into the mode sum for the
propagator (\ref{modesum}),
\begin{eqnarray}
\lefteqn{i\Delta(x;x') = } \nonumber \\
& & \hspace{-.2cm} \frac{\Bigl[ (1\!-\!\epsilon)^2 H
H'\Bigr]^{\frac{D}2-1}}{ (4 \pi)^{\frac{D}2}} \frac{\pi^{\frac32}
\, 2^{\frac{D-3}2}}{ \Bigl(\frac{\Delta x}{\sqrt{\eta
\eta'}}\Bigr)^{\frac{D-3}2}} \int_{z_0}^{\infty} \! dz \,
z^{\frac{D-1}2}
J_{\frac{D-3}2}\Bigl(\frac{\Delta x}{\sqrt{\eta \eta'}} \, z\Bigr) \nonumber \\
& & \hspace{.8cm} \times \Biggl\{\theta(\Delta \eta)
H^{(1)}_{\nu}\Bigl( \sqrt{\frac{\eta}{\eta'}} \, z \Bigr)
H^{(1)}_{\nu}\Bigl(\sqrt{\frac{\eta'}{ \eta}} \, z\Bigr)^* +
\theta(-\Delta \eta) \times \Bigl({\rm conjugate}\Bigr) \Biggr\} .
\qquad \label{newsum}
\end{eqnarray}
Here and subsequently $z_0 \equiv k_0 \sqrt{\eta \eta'}$.

We can obviously break the integral over $z$ up into two parts,
\begin{equation}
\int_{z_0}^{\infty} dz = \int_0^{\infty} dz - \int_0^{z_0} dz \; .
\label{trivial}
\end{equation}
This means that the result for (\ref{newsum}) is what we already
have (\ref{problem}) from the work of \cite{JP1, JMP,JP2}, minus
the finite range integral. It would be simple enough to expand the
integrand of this second contribution and then integrate termwise,
but we really only need the most infrared singular parts. For the
inflationary case of $0 \leq \epsilon < 1$ the index $\nu$ is
positive and the most infrared singular parts of the integrand
derive from the $J_{-\nu}$ contributions to the Hankel functions.
 For the decelerating case of $1 < \epsilon \leq 2(D\!-\!1)/D$ the
index $\nu$ is negative and it is the $J_{+\nu}$ parts of the
Hankel functions that are the most infrared singular. We shall
work out the series of leading corrections in each case.

Let us begin with the most infrared singular correction for the
inflationary case of $0 \leq \epsilon < 1$. From the small $z$ forms
(\ref{L1}) and (\ref{L2}) we see that the desired correction is,
\begin{eqnarray}
\delta i \Delta_0(x;x') &=& -\lefteqn{\frac{\Bigl[ (1 \!-\!
\epsilon)^2 H H'\Bigr]^{\frac{D}2-1} }{ (4 \pi)^{\frac{D}2}}
\,\frac{2^{2 \nu} \Gamma^2(\nu)}{\pi^{\frac12}
\Gamma(\frac{D-1}2)}\, \int_0^{z_0} \! dz \, z^{D-2-2\nu} }
\nonumber \\
&=& \frac{\Bigl[ (1\!-\!\epsilon)^2 H H'\Bigr]^{\frac{D}2-1} }{ (4
\pi)^{\frac{D}2}} \, \frac{\Gamma(2 \nu)
\Gamma(\nu)}{\Gamma(\frac12 \!+\! \nu) \Gamma(\frac{D-1}2)} \,
\frac{2 (1 \!-\! \epsilon)}{\epsilon (D \!-\!2)}
\Bigl(\frac1{k_0^2 \eta \eta'}\Bigr)^{\frac{\epsilon (D-2)}{2
(1-\epsilon)}} \; . \qquad \label{0mode}
\end{eqnarray}
To reach the last form (\ref{0mode}) we have used the doubling
formula for the Gamma function,
\begin{equation}
\Gamma(2x) = \frac{2^{2x -1}}{\pi^{\frac12}} \, \Gamma(x)
\Gamma\Bigl(x \!+\! \frac12\Bigr) \; .
\end{equation}
Of course the infrared divergence at $z=0$ in $\delta i\Delta_0$
was dimensionally regulated, the same way as in the infinite
space result $i\Delta_{\infty}$. This is wrong for $\delta i
\Delta_0$, just as it was for $i\Delta_{\infty}$, but expression
(\ref{trivial}) implies that the two errors must cancel and we
will shortly see this explicitly.

We shall use the notation $\delta i\Delta_N$ to indicate the
$N$-th order correction in the case of an inflationary
universe ($\epsilon<1$). For the correction in the decelerating
case ($1<\epsilon\leq 2(D-1)/D$) we use the notation $\delta i
\Delta^N$. Before deriving the higher corrections, let us see that
the addition of (\ref{0mode}) eliminates the $N=0$ divergence
(\ref{lower}). Both (\ref{0mode}) and the $n=0$ term from the last
line of (\ref{problem}) have a common factor that we may as well
omit,
\begin{equation}
\frac{\Bigl[ (1\!-\!\epsilon)^2 H H'\Bigr]^{\frac{D}2-1} }{(4
\pi)^{\frac{D}2}} \times \frac1{\Gamma(\frac12 \!+\! \nu)} \; .
\label{common}
\end{equation}
The remaining contributions are,
\begin{eqnarray}
\lefteqn{ \frac{2 (1 \!-\! \epsilon) \Gamma(2\nu)
\Gamma(\nu)}{\epsilon (D\!-\!2) \Gamma(\frac{D-1}2)}
\Bigl(\frac1{k_0^2 \eta \eta'}\Bigr)^{\frac{\epsilon (D-2)}{2
(1-\epsilon)}} - \frac{\Gamma(\frac{D}2 \!-\! 1) \Gamma(2 \!-\!
\frac{D}2)}{\Gamma(\frac12 \!-\! \nu)} \, \frac{\Gamma(\frac{D-1}2
\!+\! \nu) \Gamma(\frac{D-1}2 \!-\!
\nu)}{\Gamma(\frac{D}2)} } \nonumber \\
& & = \frac{2 (1 \!-\! \epsilon) \Gamma(2\nu) \Gamma(\nu)}{\epsilon
(D\!-\!2) \Gamma(\frac{D-1}2)} \Biggl\{\Bigl(\frac1{k_0^2 \eta
\eta'}\Bigr)^{\frac{
\epsilon (D-2)}{2 (1-\epsilon)}} \nonumber \\
& & \hspace{4.3cm} + \frac{\Gamma(\frac{D-1}2)}{\Gamma(\nu)}
\frac{\Gamma(1 \!-\! \frac{D}2)}{\Gamma(\frac12 \!-\! \nu)} \,
\frac{\Gamma(\frac{D-1}2 \!+\! \nu)}{\Gamma(2\nu)} \,
\frac{\Gamma(\frac{D-1}2 \!-\! \nu)}{\frac{2 (1-\epsilon)}{\epsilon
(D-2)}} \Biggr\} . \qquad \label{factors}
\end{eqnarray}
It is convenient to expand this expression in terms of the small
parameter $\alpha \equiv \epsilon (D \!-\!2)/[ 2(1 \!-\!
\epsilon)]$
such that in the limit when $\epsilon$ vanishes expression
(\ref{factors}) reduces to
\begin{eqnarray}
&&\hskip -.5cm \lim_{\epsilon \rightarrow 0} \frac{2 (1 \!-\! \epsilon)
\Gamma(2\nu) \Gamma(\nu)}{\epsilon (D\!-\!2) \Gamma(\frac{D-1}2)}
\Biggl\{\Bigl(\frac1{k_0^2 \eta \eta'}\Bigr)^{
\frac{\epsilon (D-2)}{2 (1-\epsilon)}}
+ \frac{\Gamma(\frac{D-1}2)}{\Gamma(\nu)} \frac{\Gamma(1 \!-\!
\frac{D}2)}{\Gamma(\frac12 \!-\! \nu)} \, \frac{\Gamma(\frac{D-1}2
\!+\! \nu)}{\Gamma(2\nu)} \, \frac{\Gamma(\frac{D-1}2 \!-\!
\nu)}{\frac{2 (1-\epsilon)}{\epsilon (D-2)}} \Biggr\} \hskip 1.cm
\label{final} \\
&= & \hspace{-.1cm} \Gamma(D \!-\!1) \biggl\{ \ln(a a') - \pi
\cot\Bigl(
\frac{\pi D}2\Bigr)
+ 2 \ln\Bigl(\frac{H_0}{k_0}\Bigr) +
\psi\Bigl(\frac{D\!-\!1}2\Bigr) - \psi\Bigl(\frac{D}2\Bigr) +
\psi(D \!-\! 1) - \gamma \biggr\} \, ,
\nonumber
\end{eqnarray}
where $\psi(z)=({d}/{dz})\ln(\Gamma(z))$ indicates the digamma
function, we used $a = -1/H_0\eta$ (valid for $\epsilon = 0$),
$\psi(1) = -\gamma$ and the reflection formula for the digamma
function,
\begin{equation}
\psi(1 \!-\! x) = \psi(x) + \pi \cot(\pi x) \; .
\end{equation}
Multiplying (\ref{final}) by the common factor (\ref{common}), and
then adding the rest of (\ref{problem}) --- which is not singular
for $\epsilon = 0$ --- gives the following result,
\begin{eqnarray}
\lefteqn{\lim_{\epsilon \rightarrow 0} i\Delta(x;x') =
\frac{H_0^{D-2}}{(4\pi)^{\frac{D}2}} \Biggl\{
\frac{\Gamma(\frac{D}2)}{\frac{D}2-1} \,
\Bigl(\frac{4}{y}\Bigr)^{\frac{D}2-1} \!\!+ \frac{\Gamma(\frac{D}2
\!+\!1)}{\frac{D}2 \!-\! 2} \, \Bigl(\frac{4}y\Bigr)^{\frac{D}2-2}
\!\!+ \frac{\Gamma(D\!-\!1)}{ \Gamma(\frac{D}2)} \biggl[ \ln(a a')
}
\nonumber \\
& & - \pi \cot\Bigl( \frac{\pi D}2\Bigr) + 2
\ln\Bigl(\frac{H_0}{k_0}\Bigr) + \psi\Bigl(\frac{D\!-\!1}2\Bigr) -
\psi\Bigl(\frac{D}2\Bigr) + \psi(D
\!-\! 1) - \gamma \biggr] \qquad \nonumber \\
& & + \sum_{n=1}^{\infty} \Biggl[ \frac{\Gamma(D \!-\! 1 \!+\! n)}{n
\, \Gamma(\frac{D}2 \!+\! n)} \Bigl(\frac{y}4\Bigr)^n -
\frac{\Gamma(\frac{D}2 \!+\! 1 \!+\! n)}{(2 \!-\! \frac{D}2 \!+\! n)
\, (n \!+\! 1)!} \Bigl(\frac{y}4\Bigr)^{n - \frac{D}2 +2} \Biggr] +
O(k_0^2) \Biggr\} . \qquad \label{deSitter}
\end{eqnarray}
Except for the order $k_0^2$ corrections, and for some constant,
finite factors on the second line, expression (\ref{deSitter})
agrees precisely with the result first obtained in \cite{OW} and
used subsequently in many one and two loop computations
\cite{MW,KW,BOW,PTW,many} on de Sitter background.

It is straightforward to work out the next contributions from the
lower limit. We merely add up the three first order corrections from
the Bessel and Hankel functions for the case of $\nu$ positive,
\begin{eqnarray}
J_{\frac{D-3}2}\Bigl( \frac{\Delta x}{\sqrt{\eta \eta'}} \, z\Bigr)
& = & \Bigl( \frac{\Delta x}{2 \sqrt{\eta \eta'}}
\Bigr)^{\frac{D-3}2} \times
\frac{z^{\frac{D-3}2}}{\Gamma(\frac{D-1}2)} \Biggl\{ 1 -
\frac{\frac{\Delta x^2}{\eta \eta'} \, \frac{z^2}{4}}{(\frac{D-1}2)}
+ O(z^4) \Biggr\} \; , \\
H^{(1)}_{\nu}\Bigl(\sqrt{\frac{\eta}{\eta'}} \, z\Bigr) & = &
\frac{-i (\frac12 z)^{-\nu}(\eta/\eta')^{-\nu/2}}{\sin(\nu \pi)
\Gamma(1\!-\!\nu)} \Biggl\{1 - \frac{ \frac{\eta}{\eta'} \,
\frac{z^2}{4}}{1 \!-\! \nu}
+ O(z^4) \Biggr\} \; , \\
H^{(1)}_{\nu}\Bigl(\sqrt{\frac{\eta'}{\eta}} \, z \Bigr)^* & = &
\frac{i (\frac12 z)^{-\nu}(\eta'/\eta)^{-\nu/2}}{\sin(\nu \pi)
\Gamma(1\!-\!\nu)} \Biggl\{1 - \frac{ \frac{\eta'}{\eta} \,
\frac{z^2}{4}}{1 \!-\! \nu} + O(z^4) \Biggr\} \; .
\end{eqnarray}
The resulting lower limit term is,
\begin{eqnarray}
\delta i \Delta_1=-\lefteqn{\frac{\Bigl[ (1\!-\!\epsilon)^2 H
H'\Bigr]^{\frac{D}2-1} }{(4 \pi)^{ \frac{D}2}} \frac{2 \Gamma(2
\nu) \Gamma(\nu)}{\Gamma(\frac12 \!+\! \nu) \Gamma(\frac{D-1}2)}
\Biggl[ \frac{(\eta^2 \!+\! {\eta'}^2)}{4 (\nu \!-\! 1) \eta
\eta'} - \frac{\Delta x^2}{2 (D \!-\!1) \eta \eta'} \Biggr]
\!\! \int_0^{z_0} \!\!\! dz \, z^{D-2\nu} } \nonumber \\
& & = \frac{\Bigl[ (1\!-\!\epsilon)^2 H H'\Bigr]^{\frac{D}2-1} }{(4
\pi)^{ \frac{D}2}} \frac{2 \Gamma(2 \nu) \Gamma(\nu)}{\Gamma(\frac12
\!+\! \nu) \Gamma(\frac{D-1}2)} \nonumber \\
& & \hspace{4.7cm} \times \Biggl[ \frac{(\eta^2\!+\!{\eta'}^2)}{4
(\nu \!-\! 1) \eta \eta'} - \frac{\Delta x^2}{2 (D \!-\!1) \eta
\eta'} \Biggr] \frac{- z_0^{D+1-2\nu}}{D \!+\! 1 \!-\! 2\nu} \; .
\qquad \label{term1}
\end{eqnarray}
This should cancel the $N=1$ diverge of (\ref{lower}) at $\epsilon =
2/D$, which affects the $n=0$ and $n=1$ terms on the last line of
(\ref{problem}),
\begin{equation}
\frac{\Bigl[ (1\!-\!\epsilon)^2 H H'\Bigr]^{\frac{D}2-1} }{(4
\pi)^{\frac{D}2}} \frac{\Gamma(\frac{D+1}2 \!+\! \nu)
\Gamma(\frac{D+1}2 \!-\! \nu) \Gamma(-\frac{D}2)}{\Gamma(\frac12
\!+\! \nu) \Gamma(\frac12 \!-\! \nu)}
\Biggl[\frac{-\frac{D}2}{(\frac{D-1}2)^2 \!-\! \nu^2} - \frac{y}4
\Biggr] \; . \label{term2}
\end{equation}

The key to seeing that the infrared divergence of (\ref{term1})
cancels that in (\ref{term2}) is to express both in terms of the
small parameter,
\begin{equation}
\alpha \equiv \nu - \Bigl(\frac{D\!+\!1}2\Bigr) = \frac{(D \epsilon
\!-\! 2)}{ 2 (1 \!-\! \epsilon)} \; .
\end{equation}
As before, we extract the common factor of,
\begin{equation}
\frac{\Bigl[ (1\!-\!\epsilon)^2 H H'\Bigr]^{\frac{D}2-1} }{(4
\pi)^{\frac{D}2}} \, \frac1{\Gamma(\frac12 \!+\! \nu)} =
\frac{\Bigl[ (1\!-\!\epsilon)^2 H H'\Bigr]^{\frac{D}2-1} }{(4
\pi)^{\frac{D}2}} \, \frac1{\Gamma(\frac{D}2 \!+\! 1 \!+\! \alpha)}
\; . \label{factor}
\end{equation}
The lower limit contribution from (\ref{term1}) is this factor
times,
\begin{eqnarray}
\lefteqn{\frac{2 \Gamma(2 \nu) \Gamma(\nu)}{\Gamma(\frac{D-1}2)}
\Biggl[ \frac{(\eta^2\!+\!{\eta'}^2)}{4 (\nu \!-\! 1) \eta \eta'} -
\frac{\Delta x^2}{2 (D \!-\!1) \eta \eta'} \Biggr]
\frac{- z_0^{D+1-2\nu}}{D \!+\! 1 \!-\! 2\nu} } \nonumber \\
& & \hspace{.5cm} = \frac{\Gamma(D \!+\! 1 \!+\! 2\alpha)
\Gamma(\frac{D+1}2 \!+\! \alpha)}{4 \alpha \Gamma(\frac{D+1}2)}
\Biggl[ \frac1{1 \!+\! \frac{2 \alpha}{D-1}} \Bigl(\frac{\eta^2
\!+\! {\eta'}^2}{\eta \eta'}\Bigr) - \frac{\Delta x^2}{\eta
\eta'}\Biggr] \Bigl(\frac1{k_0^2 \eta \eta'}
\Bigr)^{\alpha} \; , \qquad \\
& & \hspace{.5cm} = \frac{\Gamma(D \!+\! 1)}{4} \Biggl\{ \frac{2
\!-\! y}{ \alpha} + (2 \!-\! y) \Biggl[ 2 \psi(D\!+\!1) +
\psi\Bigl(\frac{D \!+\!1}2
\Bigr) + \ln\Bigl(\frac1{ k_0^2 \eta \eta'}\Bigr)\Biggr] \qquad \nonumber \\
& & \hspace{7cm} - \frac2{D \!-\!1} \Bigl( \frac{\eta^2 \!+\!
{\eta'}^2}{ \eta \eta'}\Bigr) + O(\alpha) \Biggr\} . \qquad
\label{fin1}
\end{eqnarray}
In contrast, the contribution from (\ref{term2}) is (\ref{factor})
times,
\begin{eqnarray}
\lefteqn{\frac{\Gamma(\frac{D+1}2 \!+\! \nu) \Gamma(\frac{D+1}2
\!-\! \nu) \Gamma(-\frac{D}2)}{\Gamma(\frac12 \!-\! \nu)}
\Biggl[\frac{-\frac{D}2}{
(\frac{D-1}2)^2 \!-\! \nu^2} - \frac{y}4 \Biggr] } \nonumber \\
& & \hspace{1cm} =-\frac{\Gamma(D \!+\! 1 \!+\! \alpha) \Gamma(1
\!-\! \alpha) \Gamma(-\frac{D}2)}{4 \alpha \Gamma(-\frac{D}2 \!-\!
\alpha)} \Biggl[
\frac2{(1 \!+\! \frac{\alpha}{D}) (1 \!+\! \alpha)} - y\Biggr] \; , \\
& & \hspace{1cm} = \frac{\Gamma(D \!+\! 1)}{4} \Biggl\{
-\Bigl(\frac{2 \!-\! y}{\alpha}\Bigr) + (2 \!-\! y) \Biggl[ -\psi(D
\!+\! 1) + \psi(1) -
\psi\Bigl(-\frac{D}2\Bigr) \Biggr] \qquad \nonumber \\
& & \hspace{7cm} + 2 \Bigl(\frac{D \!+\! 1}{D}\Bigr) +
O(\alpha)\Biggr\} . \qquad \label{fin2}
\end{eqnarray}
Adding (\ref{fin1}) to (\ref{fin2}) and taking $\alpha = 0$ (which
implies $\epsilon = 2/D$) gives,
\begin{eqnarray}
\lefteqn{ \frac{\Gamma(D \!+\! 1)}4 \Biggl\{ (2 \!-\! y) \Biggl[
\ln\Bigl(\frac1{k_0^2 \eta \eta'}\Bigr) + \psi(D \!+\! 1) +
\psi\Bigl(\frac{D \!+\! 1}2\Bigr) - \gamma -
\psi\Bigl(-\frac{D}2\Bigr) \Biggr] } \nonumber \\
& & \hspace{6cm} - \frac2{D \!-\!1} \Bigl( \frac{\eta^2 \!+\!
{\eta'}^2}{
\eta \eta'}\Bigr) + 2 \Bigl(\frac{D \!+\! 1}{D}\Bigr) \Biggr\} , \qquad \\
& & \hspace{0cm} = \frac{\Gamma(D \!+\! 1)}4 \Biggl\{ (2 \!-\! y)
\Biggl[ \Bigl(\frac{D\!-\!2}{D}\Bigr) \ln(a a') + 2
\ln\Bigl[\frac{(D \!-\! 2) H_0}{
D k_0}\Bigr] - \pi \cot\Bigl(\frac{\pi D}2\Bigr) \nonumber \\
& & \hspace{3.5cm} + \psi(D \!+\! 1) + \psi\Bigl(\frac{D \!+\!
1}2\Bigr) - \gamma - \psi\Bigl(\frac{D}2 \!+\! 1\Bigr) \Biggr] \nonumber \\
& & \hspace{6.5cm} - \frac2{D \!-\!1} \Bigl( \frac{\eta^2 \!+\!
{\eta'}^2}{\eta \eta'}\Bigr) + 2 \Bigl(\frac{D \!+\! 1}{D}\Bigr)
\Biggr\} . \qquad \label{finN1}
\end{eqnarray}

We can get the full propagator for $\epsilon = 2/D$ by multiplying
(\ref{finN1}) by the common factor (\ref{factor}), and then adding
the rest of (\ref{problem}) with the now finite $N=0$ correction
(\ref{0mode}),
\begin{eqnarray} \lefteqn{\lim_{\epsilon
\rightarrow \frac{2}{D}} i\Delta(x;x') = }
\nonumber \\
& & \frac{[(1 \!-\! \frac2{D})^2 H H']^{\frac{D}2-1}}{
(4\pi)^{\frac{D}2}} \Biggl\{
\frac{\Gamma\big(\frac{D}{2}+1\big)}{\big(1-\frac{D}{2}\big)\big(-\frac{D}{2}\big)}
\, \Bigl(\frac{4}{y}\Bigr)^{ \frac{D}2-1} + \frac{\Gamma(\frac{D}2
\!+\!2)}{(2 \!-\! \frac{D}2) (1 \!-\! \frac{D}2)} \,
\Bigl(\frac4{y}\Bigr)^{\frac{D}2-2}
\nonumber \\
& & + \frac1{2!} \frac{\Gamma(\frac{D}2 \!+\!3)}{(3 \!-\! \frac{D}2)
(2 \!-\! \frac{D}2)} \, \Bigl(\frac4{y}\Bigr)^{\frac{D}2-3} +
\frac{\Gamma(D\!+\!1)}{4 \Gamma(\frac{D}2 \!+\! 1)} \Biggl[ \frac{2
(D\!-\!1)}{k_0^2 \eta \eta'} - \frac2{D \!-\! 1}
\Bigl(\frac{\eta}{\eta'} \!+\! \frac{\eta'}{\eta}\Bigr) \nonumber \\
& & \hspace{2cm} + 2 + \frac2{D} + (2\!-\!y) \Bigl\{ \Bigl(1 \!-\!
\frac2{D}\Bigr) \ln(a a') - \pi \cot\Bigl( \frac{\pi D}2\Bigr) + K_D
\Bigr\} \Biggr] \nonumber \\
& & + \sum_{n=2}^{\infty} \Biggl[ \frac{\Gamma(\frac{D}2 \!+\! 2
\!+\! n)}{(2 \!-\! \frac{D}2 \!+\! n) (1 \!-\! \frac{D}2 \!+\! n) \,
(n \!+\! 1)!} \Bigl(\frac{y}4\Bigr)^{n - \frac{D}2 +2} \nonumber \\
& & \hspace{5.5cm} - \frac{\Gamma(D \!+\! n)}{n (n \!-\!1) \,
\Gamma(\frac{D}2 \!+\! n)} \Bigl(\frac{y}4\Bigr)^n \Biggr] +
O(k_0^2) \Biggr\} . \qquad \label{N=1}
\end{eqnarray}
Here the constant $K_D$ is,
\begin{equation}
K_D \equiv 2 \ln\Bigl[\Bigl(1 \!-\! \frac2{D}\Bigr)
\frac{H_0}{k_0}\Bigr] + \psi(D \!+\! 1) + \psi\Bigl(\frac{D \!+\!
1}2\Bigr) - \gamma - \psi\Bigl(\frac{D}2 \!+\! 1\Bigr) \; .
\end{equation}
As far as we know the literature contains no result against which we
can check (\ref{N=1}) but its limit in $D=4$ dimensions has been
worked out,
\begin{eqnarray}
\lefteqn{\lim_{D \rightarrow 4} \lim_{\epsilon \rightarrow
\frac{2}{D}} i\Delta(x;x') = \frac{H H'}{64 \pi^2} \Biggl\{
\frac4{y} + \frac{18}{k_0^2 \eta \eta'} - 2 \Bigl(\frac{\eta}{\eta'}
\!+\! \frac{\eta'}{\eta}\Bigr) - 11 y + 16 } \nonumber \\
& & \hspace{2cm} + 3 (2 \!-\! y) \Bigl[-\ln\bigl(y\bigr) + 2
\ln\Bigl(\frac{H_0}{2 k_0}\Bigr) + \frac12 \ln(a a') -2\gamma
\Bigr] + O(k_0^2) \Biggl\} . \qquad
\end{eqnarray}
This agrees perfectly with equation (3.82) of \cite{ITTW}.

We have seen that the lower limit term which corrects the $N=0$
problem in (\ref{lower}) is given by (\ref{0mode}).
For the $N=1$ problem the corresponding lower limit correction is
(\ref{term1}).
To see the general pattern, first substitute the relation for $H$ in
terms of $\eta$,
\begin{equation}
H = H_0 \Bigl[ -(1\!-\!\epsilon) H_0 \eta\Bigr]^{\frac{\epsilon}{1
-\epsilon}} \; .
\end{equation}
This reveals the $N=0$ correction (\ref{0mode}) {\it to be
constant},
\begin{equation}
\delta i\Delta_0 \equiv \frac{\Bigl[ (1\!-\!\epsilon)^2 H_0^2
\Bigr]^{\frac{D}2 -1}}{(4\pi)^{\frac{D}2}} \frac{\Gamma(2 \nu)
\Gamma(\nu)}{\Gamma(\frac12 \!+\! \nu) \Gamma(\frac{D-1}2)} \,
\frac{2 (1\!-\!\epsilon)}{\epsilon (D \!-\! 2)} \Biggl[\frac{(1
\!-\! \epsilon)^2 H_0^2}{k_0^2}\Biggr]^{\frac{\epsilon (D-2)}{2 (1 -
\epsilon)}} \; .
\end{equation}
The same substitution reveals that the $N=1$ correction
(\ref{term1}) is quadratic,
\begin{eqnarray}
\lefteqn{\delta i\Delta_1 \equiv \frac{\Bigl[ (1\!-\!\epsilon)^2
H_0^2 \Bigr]^{\frac{D}2 -1}}{(4\pi)^{\frac{D}2}} \frac{\Gamma(2 \nu)
\Gamma(\nu)}{
\Gamma(\frac12 \!+\! \nu) \Gamma(\frac{D-1}2)} } \nonumber \\
& & \hspace{2cm} \times \Biggl[ \frac{k_0^2 (\eta^2 \!+\!
{\eta'}^2)}{4 (\nu \!-\! 1)} - \frac{k_0^2 \Delta x^2}{2 (D \!-\!
1)} \Biggr] \, \frac{2 (1 \!-\! \epsilon)}{(D \epsilon \!-\! 2)}
\Biggl[\frac{(1 \!-\! \epsilon)^2 H_0^2}{k_0^2}
\Biggr]^{\frac{\epsilon (D-2)}{ 2 (1-\epsilon)}} . \qquad
\end{eqnarray}
Both corrections are homogeneous solutions of the propagator
equation (\ref{propeqn}),
\begin{equation}
\partial_{\mu} \Bigl(\sqrt{-g} g^{\mu\nu} \partial_{\nu} \delta i\Delta_N\Bigr)
= 0 \; . \label{homo}
\end{equation}
Note that each lower limit correction $\delta i\Delta_N$ must
separately solve (\ref{homo}) because each goes like a distinct
power of $k_0$. {\it The freedom to add such homogeneous terms is
precisely what is not fixed by just solving the propagator equation
rather than using the mode sum.}

We could work out the N-th lower limit correction $\delta i\Delta_N$
from the mode sum but that would involve tedious multiplications of
corrections from the Bessel function and the two Hankel functions. A
simpler technique is to use the fact that the correction must have
the form,
\begin{eqnarray}
\lefteqn{\delta i\Delta_N = \frac{\Bigl[ (1\!-\!\epsilon)^2 H_0^2
\Bigr]^{\frac{D}2 -1}}{(4\pi)^{\frac{D}2}} \frac{\Gamma(2 \nu)
\Gamma(\nu)}{\Gamma(\frac12 \!+\! \nu) \Gamma(\frac{D-1}2)} \,
\frac{2 (1\!-\!\epsilon)}{(D\!-\!2\!+\!2N) \epsilon \!-\! 2 N} }
\nonumber \\
& & \hspace{2.5cm} \times \Biggl[\frac{(1 \!-\! \epsilon)^2
H_0^2}{k_0^2} \Biggr]^{\frac{\epsilon (D-2)}{2 (1 - \epsilon)}}
k_0^{2N} \sum_{k=0}^{N} \sum_{\ell = 0}^{N-k} a_{k\ell} \Delta
x^{2k} \eta^{2\ell} {\eta'}^{2(N-k-\ell)} \; . \qquad \label{ansatz}
\end{eqnarray}
Then we determine the coefficients $a_{k\ell}$ by three
requirements:
\begin{enumerate}
\item{The coefficient $a_{N0}$ derives entirely from the $z^{2N}$ correction
of the Bessel function and, by direct examination of
(\ref{newsum}), we can see that it is,
\begin{equation}
a_{N0} = \frac{(-1)^N \Gamma(\frac{D-1}2)}{N! 4^N \Gamma(\frac{D-1}2
\!+\! N)} \; ; \label{akk}
\end{equation}}
\item{Symmetry under $\eta \leftrightarrow \eta'$ implies,
\begin{equation}
a_{k\ell} = a_{k (N-k-\ell)} \; ; \label{reflect}
\end{equation}}
\item{The series must of course solve (\ref{homo}).}
\end{enumerate}

The differential equation (\ref{homo}) implies,
\begin{eqnarray}
0 & = & \Bigl[\partial^2 + \frac{2 \nu \!-\!1}{\eta}\,
\partial_0\Bigr] \sum_{k=0}^{N} \sum_{\ell=0}^{N-k}
a_{k\ell} \Delta x^{2k} \eta^{2\ell} {\eta'}^{2(N-k-\ell)} \; , \\
& = & \sum_{k=0}^{N} \sum_{\ell=0}^{N-k} 2k (2k \!+\! D \!-\! 3)
a_{k\ell}
\Delta x^{2k-2} \eta^{2\ell} {\eta'}^{2(N-k-\ell)} \nonumber \\
& & \hspace{.5cm} - \sum_{k=0}^{N} \sum_{\ell=0}^{N-k} 4\ell
(\ell-\nu) a_{k\ell}
\Delta x^{2k} \eta^{2\ell-2} {\eta'}^{2(N-k-\ell)} \; , \\
& = & \sum_{k=0}^{N-1} \sum_{\ell=0}^{N-1-k} 2(k\!+\!1) (2k \!+\! D
\!-\! 1)
a_{k+1 \,\ell} \Delta x^{2k} \eta^{2\ell} {\eta'}^{2(N-1-k-\ell)} \nonumber \\
& & \hspace{.5cm} - \sum_{k=0}^{N-1} \sum_{\ell=0}^{N-1-k}
4(\ell\!+\!1) (\ell+1-\nu) a_{k\, \ell+1} \Delta x^{2k}
\eta^{2\ell} {\eta'}^{2(N-1-k-\ell)} \!. \qquad
\end{eqnarray}
Hence the coefficients must obey,
\begin{equation}
(k\!+\!1) (2k \!+\! D \!-\! 1) a_{k+1 \, \ell} = 2(\ell \!+\! 1)
(\ell \!+\! 1 \!-\!  \nu ) a_{k \, \ell+1} \; . \label{recursion}
\end{equation}
The unique solution consistent with the other two of the three
properties is,
\begin{equation}
a_{k\ell} = \Bigl(-\frac1{4}\Bigr)^N  \frac1{k! \, \ell! \, (N
\!-\! k \!-\! \ell)!} \, \frac{\Gamma(\frac{D-1}2) \, \Gamma^2(1
\!-\! \nu)}{ \Gamma(k \!+\! \frac{D-1}2) \Gamma(\ell \!+\! 1 \!-\!
\nu) \Gamma(N \!-\! k \!-\! \ell \!+\! 1 \!-\! \nu)} \; .
\label{akl}
\end{equation}
For $N=0$ this gives the known result ,
\begin{equation}
N=0 \qquad \Longrightarrow \qquad a_{00} = 1 \; .
\end{equation}
A less trivial check is that it also works for $N=1$,
\begin{equation}
N=1 \qquad \Longrightarrow \qquad a_{00} = a_{01} = \frac1{4 (\nu
\!-\! 1)} \qquad {\rm and} \qquad a_{10} = -\frac1{2(D \!-\! 1)} \;
.
\end{equation}

Let us turn now to the decelerating case of $1 < \epsilon \leq
2(D\!-\!1)/D$ for which the infinite space propagator
(\ref{problem}) diverges at the discrete values given in
(\ref{upper}). By paralleling what we did for the inflationary
case one can show that the lower limit term which corrects the
$N=0$ problem is,
\begin{eqnarray}
\delta i\Delta^0 &=& -\frac{\Bigl[ (1\!-\!\epsilon)^2 H H'
\Bigr]^{\frac{D}2 -1}}{(4\pi)^{\frac{D}2}} \frac{\Gamma(-2 \nu)
\Gamma(-\nu)}{ \Gamma(\frac12 \!-\! \nu) \Gamma(\frac{D-1}2)}
\,2 \int_0^{z_0} \!\! dz \, z^{D-2+2\nu} \; ,  \nonumber \\
&=& \frac{\Bigl[ (1\!-\!\epsilon)^2 H H'\Bigr]^{\frac{D}2 -1}}{
(4\pi)^{\frac{D}2}} \frac{\Gamma(-2 \nu)
\Gamma(-\nu)}{\Gamma(\frac12\!-\!\nu) \Gamma(\frac{D-1}2)} \,
\frac{2 (\epsilon \!-\!1)}{2(D \!-\! 1) \!-\! D \epsilon}
\Bigl(\frac1{k_0^2 \eta \eta'}\Bigr)^{\frac{2 (D-1) - D\epsilon}{2
(\epsilon -1)}} , \qquad \\
&=& \frac{\Bigl[ (1\!-\!\epsilon)^2 H_0^2\Bigr]^{\frac{D}2 -1}}{
(4\pi)^{\frac{D}2}} \frac{\Gamma(-2 \nu)
\Gamma(-\nu)}{\Gamma(\frac12\!-\!\nu) \Gamma(\frac{D-1}2)} \,
\frac{2 (\epsilon \!-\!1)}{2(D \!-\! 1) \!-\! D \epsilon}
\Biggl[\frac{(1 \!-\! \epsilon)^2 H_0^2}{k_0^2}
\Biggr]^{\frac{\epsilon (D-2)}{2 (1 - \epsilon)}} \Bigl(k_0^2 \eta
\eta'\Bigr)^{2 \nu} \; . \qquad
\end{eqnarray}

One can easily check that $(k_0^2 \eta \eta')^{2\nu}$ solves the
homogeneous equation (\ref{homo}). So the full series of these lower
limit corrections should take the form,
\begin{eqnarray}
\lefteqn{\delta i\Delta^N = \frac{\Bigl[ (1\!-\!\epsilon)^2 H_0^2
\Bigr]^{\frac{D}2 -1}}{(4\pi)^{\frac{D}2}} \frac{\Gamma(-2 \nu)
\Gamma(-\nu)}{\Gamma(\frac12\!-\!\nu) \Gamma(\frac{D-1}2)} \,
\frac{2 (\epsilon \!-\! 1)}{2 (D\!-\!1\!+\!N) \!-\! (D \!+\!
2N)\epsilon} }
\nonumber \\
& & \times \Biggl[\frac{(1 \!-\! \epsilon)^2 H_0^2}{k_0^2}
\Biggr]^{\frac{\epsilon (D-2)}{2 (1 - \epsilon)}} (k_0^2 \eta
\eta')^{2\nu} k_0^{2N} \sum_{k=0}^{N} \sum_{\ell = 0}^{N-k}
b_{k\ell} \Delta x^{2k} \eta^{2\ell} {\eta'}^{2(N-k-\ell)} \; .
\qquad \label{bansatz}
\end{eqnarray}
We determine the coefficients $b_{k\ell}$ by the same three
requirements as the $a_{k\ell}$, although the solution will be
different because the ansatz (\ref{bansatz}) is.

We need to commute the differential operator in (\ref{homo}) through
the prefactor of $(k_0^2 \eta \eta')^{2\nu}$ in the ansatz
(\ref{bansatz}),
\begin{equation}
\Bigl[\partial^2 + \frac{2\nu \!-\! 1}{\eta}\,
\partial_0 \Bigr] \Bigl(k_0^2 \eta \eta'\Bigr)^{2 \nu}
 = \Bigl( k_0^2 \eta \eta'\Bigr)^{2\nu} \Bigl[\partial^2 +
\frac{-2\nu \!-\! 1}{\eta} \, \partial_0 \Bigr] \, \; .
\end{equation}
That is a highly significant result because it means the equation
the bare series obeys is the same as we already solved for the lower
series but with the replacement $\nu \rightarrow -\nu$. So we can
write down the answer immediately,
\begin{equation}
b_{k\ell} = \Bigl(-\frac1{4}\Bigr)^N \,\frac1{k! \, \ell! \, (N
\!-\! k \!-\! \ell)!} \,\frac{\Gamma(\frac{D-1}2) \, \Gamma^2(1
\!+\! \nu)}{ \Gamma(k \!+\! \frac{D-1}2) \Gamma(\ell \!+\! 1 \!+\!
\nu) \Gamma(N \!-\! k \!-\! \ell \!+\! 1 \!+\! \nu)} \; .
\label{bkl}
\end{equation}

It is worth explicitly checking that the lowest $N$ corrections
$\delta i\Delta^N$ cancel the $\epsilon$ poles in (\ref{problem})
from the upper series (\ref{upper}). From (\ref{bansatz}) and
(\ref{bkl}) we see that the $N=0$ correction is,
\begin{equation}
\delta i\Delta^0 = \frac{[(1 \!-\! \epsilon)^{2}HH']^{\frac{D}2
-1}}{(4\pi)^{D/2}} \frac{\Gamma(-\nu)
\Gamma(-2\nu)}{\Gamma(\frac{D-1}2) \Gamma(\frac12 \!-\! \nu)}
\frac{2 (\epsilon \!-\! 1)}{\Bigl[2(D \!-\! 1) \!-\! D
\epsilon\Bigr]} \Biggl( \frac1{k_0^2\eta\eta'}
\Biggr)^{\frac{2(D-1)-D\epsilon}{2(\epsilon-1)}} \; . \label{propa0}
\end{equation}
This should cancel the divergence at $\epsilon = 2(D\!-\!1)/D$ in
the $n=0$ term on the last line of (\ref{problem}),
\begin{equation}
\frac{[(1 \!-\! \epsilon)^{2}HH']^{\frac{D}2-1}}{(4\pi)^{D/2}}
\frac{\Gamma(1 \!-\! \frac{D}2) \Gamma(\frac{D-1}2 \!+\! \nu)
\Gamma(\frac{D\!-\! 1}{2} \!-\! \nu)}{\Gamma(\frac12 \!+\! \nu)
\Gamma(\frac12 \!-\! \nu)} \;. \label{propb0}
\end{equation}
The relevant small parameter is,
\begin{eqnarray}
\alpha \equiv \frac{2(D \!-\! 1) \!-\! D \epsilon}{2(\epsilon
\!-\! 1)} = -\frac{D \!-\! 1}2 - \nu \;. \label{alpha0}
\end{eqnarray}
Adding (\ref{propa0}) to (\ref{propb0}) and taking the limit that
$\alpha$ vanishes gives,
\begin{eqnarray}
\lefteqn{\lim_{\alpha \rightarrow 0} \frac{[(1 \!-\!
\epsilon)^{2}HH']^{\frac{D}{2}-1}}{(4\pi)^{D/2}}
\frac{\Gamma(\frac{D-1}{2} \!+\! \alpha) \Gamma(D \!-\! 1 \!+\!
2\alpha)}{\Gamma(\frac{D}{2} \!+\! \alpha) \Gamma(\frac{D-1}{2}) \,
\alpha} \, \Biggl\{\Biggl(\frac{1}{k_0^2 \eta \eta'}\Biggr)^{\alpha}
} \nonumber \\
& & \hspace{2cm} -\frac{\Gamma(\frac{D-1}2)}{\Gamma(\frac{D-1}2
\!+\! \alpha)} \, \frac{\Gamma(D \!-\!1 \!+\! \alpha)}{\Gamma(D
\!-\! 1 \!+ \! 2\alpha)} \, \frac{\Gamma(1 \!-\!
\frac{D}2)}{\Gamma(1 \!-\! \frac{D}{2} \!-\! \alpha)} \,
\frac{\Gamma(1 \!-\! \alpha)}{\Gamma(1)} \Biggr\} \; . \qquad \\
& & =\frac{[(1 \!-\! \frac2{D})^{2}HH']^{\frac{D}{2} -
1}}{(4\pi)^{D/2}} \frac{\Gamma(D \!-\! 1)}{\Gamma(\frac{D}2)}
\Biggl\{ 2\ln\Bigl[\Bigl(1 \!-\! \frac2{D}\Bigr)
\frac{H_0}{k_0}\Bigr] - \Bigl(1 \!-\! \frac2{D}\Bigr) \ln(aa')
\nonumber \\
& & \hspace{2cm} -\pi\cot\Bigl(\frac{D \pi}{2}\Bigr) - \gamma -
\psi\Bigl(\frac{D}{2}\Bigr) + \psi(D\!-\! 1) + \psi\Bigl(\frac{D
\!-\! 1}{2}\Bigr) \Biggr\} \; . \qquad \label{propN0}
\end{eqnarray}
By using $\epsilon \!-\! 1 = 1 \!-\! 2/D$ the final result for the
propagator can be expressed in a form that is identical with the de
Sitter case (\ref{deSitter}),
\begin{eqnarray}
\lefteqn{\lim_{\epsilon \rightarrow \frac{2(D-1)}{D}} i\Delta(x;x')
= } \nonumber \\
& & \hspace{-.7cm} \frac{[(1 \!-\! \epsilon)^2 H H']^{\frac{D}2 -
1}}{ (4\pi)^{\frac{D}2}} \Biggl\{
\frac{\Gamma(\frac{D}2)}{\frac{D}2-1} \,
\Bigl(\frac{4}{y}\Bigr)^{\frac{D}2-1} \!\!\!\!+
\frac{\Gamma(\frac{D}2 \!+\!1)}{\frac{D}2 \!-\! 2} \,
\Bigl(\frac{4}y\Bigr)^{\frac{D}2-2} \!\!\!\!+
\frac{\Gamma(D\!-\!1)}{
\Gamma(\frac{D}2)} \Biggl[ (1 \!-\! \epsilon) \ln(a a') \nonumber \\
& & \hspace{0cm} - \pi \cot\Bigl( \frac{\pi D}2\Bigr) \!+\! 2
\ln\Bigl[|1-\epsilon| \frac{H_0}{k_0}\Bigr] \!+\!
\psi\Bigl(\frac{D\!-\!1}2\Bigr) \!-\! \psi\Bigl(\frac{D}2\Bigr)
\!+\! \psi(D \!-\! 1) \!-\! \gamma \Biggr] \qquad \nonumber \\
& & \hspace{0cm} + \sum_{n=1}^{\infty} \Biggl[ \frac{\Gamma(D \!-\!
1 \!+\! n)}{n \, \Gamma(\frac{D}2 \!+\! n)} \Bigl(\frac{y}4\Bigr)^n
- \frac{\Gamma(\frac{D}2 \!+\! 1 \!+\! n)}{(2 \!-\! \frac{D}2 \!+\!
n) \, (n \!+\! 1)!} \Bigl(\frac{y}4\Bigr)^{n - \frac{D}2 +2} \Biggr]
+ O(k_0^2) \Biggr\} . \qquad \label{e=3/2}
\end{eqnarray}
Of course one must remember that in de Sitter $\epsilon = 0$ and $H
= H_0$, so (\ref{deSitter}) and (\ref{e=3/2}) are only formally the
same.

We will content ourselves with working out one more propagator.
From (\ref{bansatz}) and (\ref{bkl}) we see that the $N=1$
correction is,
\begin{equation}
\frac{[(1 \!-\! \epsilon)^{2}HH']^{\frac{D}{2}-1}}{(4\pi)^{D/2}}
\frac{2 \Gamma(-\nu) \Gamma(-2\nu)}{\Gamma(\frac12 \!-\!
\nu)\Gamma(\frac{D-1}2)} \Bigl(\frac{-1}{4\eta\eta'} \Bigr)
\Biggl[\frac{2 \Delta x^2}{D \!-\! 1} + \frac{\eta^2 \!+\!
\eta'^2}{1 \!+\! \nu}\Biggr] \frac{(k_0^2 \eta
\eta')^{-\alpha}}{2\alpha} \; , \label{propa1}
\end{equation}
where we define the small parameter $\alpha$ as,
\begin{eqnarray}
\alpha \equiv \frac{2D \!-\! (D \!+\! 2) \epsilon}{2(\epsilon
\!-\! 1)} = -\frac{D \!+\! 1}2 - \nu \;. \label{alpha1}
\end{eqnarray}
This should cancel the divergences from the $n=0$ and $n=1$ terms on
the last line of (\ref{problem}),
\begin{equation}
\frac{[(1 \!-\! \epsilon)^{2} H H']^{\frac{D}{2}-1}}{(4\pi)^{D/2}}
\frac{\Gamma(\frac{D+1}2 \!+\! \nu) \Gamma(\frac{D+1}2 \!-\! \nu)
\Gamma(-\frac{D}2)}{\Gamma(\frac12 \!+\! \nu) \Gamma(\frac12 \!-\!
\nu)} \Biggl\{ -\frac{y}4 - \frac{\frac{D}2}{ (\frac{D-1}2)^2
\!-\! \nu^2} \Biggr\} \; . \label{propb1}
\end{equation}
Adding (\ref{propa1}) to (\ref{propb1}) and taking $\alpha$ to zero
gives,
\begin{eqnarray}
\lefteqn{ \frac{[(1 \!-\! \epsilon)^2 H H']^{\frac{D}2 -
1}}{(4\pi)^{\frac{D}2}} \, \frac{\Gamma(D \!+\! 1)}{4
\Gamma(\frac{D}2 \!+\! 1)} \Biggl\{ -\frac2{D \!-\! 1}
\Bigl(\frac{\eta}{\eta'} + \frac{\eta'}{\eta}\Bigr) + 2 + \frac2{D}
} \nonumber \\
& & \hspace{1cm} + (2 \!-\! y) \Biggl[- \Bigl(\frac{D \!-\! 2}{D
\!+\! 2}\Bigr) \ln(a a') + 2 \ln\Bigl[\Bigl(\frac{D \!-\! 2}{D \!+\!
2}\Bigr) \frac{H_0}{k_0}\Bigr] -\pi\cot\Bigl(\frac{D \pi}2\Bigr)
\nonumber \\
& & \hspace{4cm} - \gamma - \psi\Bigl(\frac{D}2 \!+\! 1\Bigr) +
\psi\Bigl(\frac{D \!+\! 1}2\Bigr) + \psi(D \!+\! 1) \Biggr] \Biggr\}
\; . \qquad
\end{eqnarray}
By taking advantage of the fact that $(1 \!-\! \epsilon) = -(D \!-\!
2)/(D \!+\! 2)$ we can express the full propagator in a form
identical to the $N=1$ result (\ref{N=1}) from the lower series,
\begin{eqnarray}
\lefteqn{\lim_{\epsilon \rightarrow \frac{2D}{D+2}} i\Delta(x;x')
} \nonumber \\
&=& \frac{[(1 \!-\! \epsilon)^2 H H']^{\frac{D}2-1}}{
(4\pi)^{\frac{D}2}} \Biggl\{ \frac{\Gamma(\frac{D}2 \!+\! 1)}{(1
\!-\! \frac{D}2) (-\frac{D}2)} \, \Bigl(\frac{4}{y}\Bigr)^{
\frac{D}2-1} + \frac{\Gamma(\frac{D}2 \!+\!2)}{(2 \!-\! \frac{D}2)
(1 \!-\! \frac{D}2)} \, \Bigl(\frac4{y}\Bigr)^{\frac{D}2-2}
\nonumber \\
& & + \frac1{2!} \frac{\Gamma(\frac{D}2 \!+\!3)}{(3 \!-\! \frac{D}2)
(2 \!-\! \frac{D}2)} \, \Bigl(\frac4{y}\Bigr)^{\frac{D}2-3} +
\frac{\Gamma(D\!+\!1)}{4 \Gamma(\frac{D}2 \!+\! 1)} \Biggl[ \frac{2
(D\!-\!1)}{k_0^2 \eta \eta'} - \frac2{D \!-\! 1}
\Bigl(\frac{\eta}{\eta'} \!+\! \frac{\eta'}{\eta}\Bigr) \nonumber \\
& & \hspace{2cm} + 2 + \frac2{D} + (2\!-\!y) \Bigl\{(1 \!-\!
\epsilon) \ln(a a') - \pi \cot\Bigl( \frac{\pi D}2\Bigr) + C_D
\Bigr\} \Biggr] \nonumber \\
& & + \sum_{n=2}^{\infty} \Biggl[ \frac{\Gamma(\frac{D}2 \!+\! 2
\!+\! n)}{(2 \!-\! \frac{D}2 \!+\! n) (1 \!-\! \frac{D}2 \!+\! n) \,
(n \!+\! 1)!} \Bigl(\frac{y}4\Bigr)^{n - \frac{D}2 +2} \nonumber \\
& & \hspace{5.5cm} - \frac{\Gamma(D \!+\! n)}{n (n \!-\!1) \,
\Gamma(\frac{D}2 \!+\! n)} \Bigl(\frac{y}4\Bigr)^n \Biggr] +
O(k_0^2) \Biggr\} . \qquad \label{Nb=1}
\end{eqnarray}
Here the constant $C_D$ is,
\begin{equation}
C_D \equiv 2 \ln\Bigl[|1 \!-\! \epsilon| \frac{H_0}{k_0}\Bigr] +
\psi(D \!+\! 1) + \psi\Bigl(\frac{D \!+\! 1}2\Bigr) - \gamma -
\psi\Bigl(\frac{D}2 \!+\! 1\Bigr) \; .
\end{equation}

\section{The scalar stress-energy tensor}

In this section we shall calculate the expectation value of the
scalar stress-energy tensor using the propagator obtained in the
previous section. The stress-energy tensor for a scalar field
$\varphi$, with the Lagrangian (\ref{Lag}), is given by
\begin{equation}
    \begin{split}
    T_{\mu\nu}&\equiv-\frac{2}{\sqrt{-g}}\frac{\delta S}{\delta g^{\mu\nu}}
=\partial_\mu\varphi\partial_\nu\varphi-\frac{1}{2} g_{\mu\nu}
    g^{\alpha\beta}\partial_\alpha\varphi\partial_\beta\varphi.
    \end{split}
\end{equation}
The expectation value with respect to the vacuum state
$|\Omega\rangle$ can be written as
\begin{equation}\label{Tmn}
     \langle
    \Omega|T_{\mu\nu}|\Omega\rangle
  =\Big(\delta^\rho_\mu\delta^\sigma_\nu
 -\frac{1}{2}g_{\mu\nu}g^{\rho\sigma}\Big)\langle
    \Omega|\partial_\rho\varphi\partial_\sigma\varphi|\Omega\rangle
    =\Big(\delta^\rho_\mu\delta^\sigma_\nu-\frac{1}{2}g_{\mu\nu}g^{\rho\sigma}\Big)\partial_\rho\partial_\sigma'i \Delta(x;x')\Big|_{x=x'},
\end{equation}
where $\partial_\mu'\equiv \frac{\partial}{\partial x'^\mu}$. The
propagator $i \Delta$ is given in terms of the infinite space
propagator (\ref{Tomislav}) and the corrections (\ref{ansatz}) and
(\ref{bansatz}),
\begin{equation}
    i \Delta(x;x')=i \Delta_\infty(x;x')+\sum_{N=0}^\infty
     \delta i\Delta_N(x;x')+\sum_{N=0}^\infty
     \delta i\Delta^N(x;x')
\,.
\end{equation}

\subsection{The infinite space contribution}

We first consider the contribution to (\ref{Tmn}) coming
from $i \Delta_\infty$. From (\ref{ydef}) we find that at
coincidence ($y\rightarrow 0$) the following two identities hold
\begin{equation}\label{identities}
    \begin{split}
        \partial_\rho y \Big|_{y=0}&=0\\
        \partial_\rho\partial_\sigma'
        y\Big|_{y=0}&=-\frac{2}{\eta^2}\eta_{\rho\sigma}
             = -2(1-\epsilon)^2H^2g_{\rho\sigma}
\,.
    \end{split}
\end{equation}
Moreover, since in dimensional regularization all $D$ dependent
powers of $y$ can be automatically subtracted, we find using
(6.131.2) in \cite{GR} that the contributions from the
hypergeometric function appearing in $i\Delta_\infty$ relevant for
this calculation are
\begin{equation}
    \begin{split}
        {}_2F_1\Big(\frac{D-1}{2}+\nu,\frac{D-1}{2}-\nu;\frac D2;1-\frac y4\Big)\Big|_{y=0}&=\frac{\Gamma(1-\frac D2)\Gamma(\frac D2)}{\Gamma(\frac 12+\nu)\Gamma(\frac 12-\nu)}\\
        \frac{d}{dy}\;{}_2F_1\Big(\frac{D-1}{2}+\nu,\frac{D-1}{2}-\nu; \frac D2;1-\frac y4\Big)\Big|_{y=0}
&=\!\!-\frac{1}{2D}\Big(\nu^2-\Big(\frac{D\!-\!1}{2}\Big)^2\Big)\frac{\Gamma(1-\frac D2)\Gamma(\frac D2)}{\Gamma(\frac 12+\nu)\Gamma(\frac 12-\nu)}.
    \end{split}
\end{equation}
Using these identities we immediately find that
\begin{equation}\label{deriv1}
    \begin{split}
        \partial_\rho\partial_\sigma'i \Delta_\infty(x;x')\Big|_{x=x'}=&H^D|1-\epsilon|^D\frac{\Gamma(1-\frac D2)}{(4\pi)^{\frac D2}}\frac{\Gamma(\frac{D-1}{2}+\nu)\Gamma(\frac{D-1}{2}-\nu)}{\Gamma(\frac{1}{2}+\nu)\Gamma(\frac{1}{2}-\nu)}\\
        &\times\bigg[\Big(\frac{D-1}{2}-\nu\Big)^2a^2\delta_\rho^0\delta_\sigma^0+\frac{1}{D}\Big(\nu^2-\Big(\frac{D-1}{2}\Big)^2\Big)g_{\rho\sigma}\bigg].
    \end{split}
\end{equation}
Making use of Eq.~(\ref{Tmn}) the one-loop contribution
to the stress-energy from  $i\Delta_\infty$ can be written as,
\begin{eqnarray}
\langle\Omega|T_{\mu\nu}|\Omega\rangle_{\infty}
  &=& \frac{H^D|1-\epsilon|^D}{(4\pi)^{D/2}}
\frac{\Gamma(1-\frac D2)\Gamma(\frac{D-1}{2}+\nu)\Gamma(\frac{D-1}{2}-\nu)}{\Gamma(\frac{1}{2}+\nu)\Gamma(\frac{1}{2}-\nu)}
\nonumber\\
        &&\times\Big(\frac{D-1}{2}-\nu\Big)\Bigg[\Big(\frac{D-1}{2}-\nu\Big)a^2\delta_\mu^0\delta_\nu^0+\frac{1}{D}\Big(\frac{(D-1)^2}{2}-\nu\Big)g_{\mu\nu}\Bigg]
\,.
\label{Tmn:infty}
\end{eqnarray}

\subsection{The $\epsilon<1$ correction}

Next we consider the $i \delta\Delta_N$ contribution
(\ref{ansatz}). We define for convenience
\begin{equation}\label{AN}
    A_N=\frac{\Big(H_0^2(1-\epsilon)^2\Big)^{\frac{D}{2}-1}}{(4\pi)^{\frac D2}}\Bigg(\frac{H_0^2(1-\epsilon)^2}{k_0^2}\Bigg)^{\frac{(D-2)\epsilon}{2(1-\epsilon)}}\frac{\Gamma(2\nu)\Gamma(\nu)}{\Gamma(\frac
    12+\nu)\Gamma(\frac{D-1}{2})}\frac{-k_0^{2N}}{N+\frac{D-1}{2}-\nu}\,
    ,
\end{equation}
such that
\begin{equation}
     \delta i\Delta_N(x;x')=A_N\sum_{k=0}^{N}\sum_{\ell=0}^{N-k}
    a_{k\ell}(\Delta x)^{2k}\eta^{2\ell}\eta'^{2(N-k-\ell)}\,.
\end{equation}
 Since at coincidence $\Delta x$ is zero, we find that the
only nonzero contribution to (\ref{Tmn}) arises from $k=1$
when both derivatives hit $\Delta x$, and from $k=0$ when both
derivatives hit $\eta$ and $\eta'$. Thus at coincidence we have
\begin{equation}
    \begin{split}
    \partial_\rho\partial_\sigma' \delta i\Delta_N(x;x')\Big|_{x=x'}=&A_N\Bigg(\sum_{\ell=0}^{N-1}a_{1\ell}\Big(-2\bar{\eta}_{\rho\sigma}\eta^{2(N-1)}\Big)\\
    &\qquad+\sum_{\ell=0}^N
    a_{0\ell}\Big(4\ell(N-\ell)\delta_\rho^0\delta_\sigma^0\eta^{2(N-1)}\Big)\Bigg)\,,
    \end{split}
\end{equation}
where
\begin{equation}
    \bar{\eta}_{\rho\sigma}=\eta_{\rho\sigma}+\delta_\rho^0\delta_\sigma^0\,.
\end{equation}
Using the expression for $a_{k\ell}$ (\ref{akl}) we find that we
can resum the series over $\ell$ to obtain
\begin{equation}\label{deriv2}
    \begin{split}
\partial_\rho\partial_\sigma' \delta i\Delta_N(x;x')\Big|_{x=x'}=&A_N \frac{(-1)^N}{\sqrt{\pi}4^\nu}\frac{\Gamma^2(1-\nu)\Gamma(N-\frac 1
2-\nu)}{\Gamma(N)\Gamma(N-\nu)\Gamma(N-2\nu)}\eta^{2(N-1)}\\
&\times\bigg(-\frac{1}{D-1}\bar{\eta}_{\rho\sigma}+\frac{N-1}{N-2\nu}\delta_\rho^0\delta_\sigma^0\bigg)\,,
\end{split}
\end{equation}
where we made use of the identity,
\begin{equation}
 \sum_{\ell = 0 }^N \frac{N!}{\ell!(N-\ell)!}\,
                      \frac{\Gamma(N\!+\!1\!-\!2\nu)}
                  {\Gamma(\ell\!+\!1-\!\nu)\Gamma(N\!+\!1\!-\!\ell\!-\!\nu)}\,
               =      \frac{\Gamma(2N\!+\!1\!-\!2\nu)}
                  {\Gamma^2(N\!+\!1-\!\nu)}\,
\,.
\label{sum:ell}
\end{equation}
 From Eqs.~(\ref{Tmn}), (\ref{AN}) and~(\ref{deriv2})
we find the following contribution to the
stress-energy tensor from the $N$-th correction,
\begin{eqnarray}
\langle\Omega|T_{\mu\nu}|\Omega\rangle_{N}
  &=& -\frac{H^D|1-\epsilon|^D}{2(4\pi)^{D/2}}
        \frac{(z_0^2)^{N+\frac{D-1}{2}-\nu}}{N+\frac{D-1}{2}-\nu}
\frac{\Gamma(1-N+\nu)\Gamma(-N+2\nu)}
     {\Gamma(\frac{3}{2}-N+\nu)\Gamma(\frac{D+1}{2})\Gamma(N)}
\label{Tmn:N<}
\\
&&\hskip -2cm
 \times\Big[\Big(\!-\!(D\!-\!1)\!+\!2\nu\!+\!(D\!-\!2)N\Big)
              a^2\delta_\mu^0\delta_\nu^0
 +\frac12 \Big(\!-\!(D\!-\!1)\!-\!2(D\!-\!3)\nu\!+\!2(D\!-\!2)N\Big)g_{\mu\nu}
   \Big]
\,,
\nonumber
\end{eqnarray}
where we have transformed the Gamma functions such that they are not singular
at the $N$-th pole~(\ref{poles:upperN}) and we used
\begin{equation}
\Big(H_0^2(1-\epsilon)^2\Big)^{\frac{D}{2}-1}\Bigg(\frac{H_0^2(1-\epsilon)^2}
{k_0^2}\Bigg)^{\frac{(D-2)\epsilon}{2(1-\epsilon)}}\frac{1}{\eta^2}
 =H^D|1-\epsilon|^D
(z_0^2)^{\frac{D-1}{2}-\nu}a^2
\,,
\end{equation}
with $z_0=k_0 \vert\eta \vert$ at coincidence.

\subsection{The $\epsilon>1$ correction}

Finally we consider the correction due to $\delta i\Delta^N$
given by Eq.~(\ref{bansatz}). We define for convenience
\begin{equation}
    B_N=\frac{\Big(H_0^2(1-\epsilon)^2\Big)^{\frac{D}{2}-1}}{(4\pi)^{\frac D2}}\Bigg(\frac{H_0^2(1-\epsilon)^2}{k_0^2}\Bigg)^{\frac{(D-2)\epsilon}{2(1-\epsilon)}}\frac{\Gamma(-2\nu)\Gamma(-\nu)}{\Gamma(\frac
    12-\nu)\Gamma(\frac{D-1}{2})}\frac{-k_0^{2N}}{N+\frac{D-1}{2}+\nu}\,
    ,
\end{equation}
such that
\begin{equation}
     \delta i\Delta^N(x;x')=B_N(k_0^2\eta\eta')^{2\nu}\sum_{k=0}^{N}\sum_{\ell=0}^{N-k}
    b_{k\ell}(\Delta x)^{2k}\eta^{2\ell}\eta'^{2(N-k-\ell)}.
\end{equation}
The calculation is similar as in the $\epsilon<1$ case, and it yields:
\begin{equation}
    \begin{split}
    \partial_\rho\partial_\sigma' \delta i\Delta^N(x;x')\Big|_{x=x'}=B_N(k_0^2\eta^2)^{2\nu}\Bigg(&\sum_{\ell=0}^{N-1}b_{1\ell}\Big(-2\bar{\eta}_{\rho\sigma}\eta^{2(N-1)}\Big)\\
    &+\sum_{\ell=0}^N
    b_{0\ell}\Big(4(\ell+\nu)(N-\ell+\nu)\delta_\rho^0\delta_\sigma^0\eta^{2(N-1)}\Big)\Bigg).
    \end{split}
\end{equation}
Summing over $\ell$ gives
\begin{equation}\label{deriv3}
    \begin{split}
\partial_\rho\partial_\sigma' \delta i\Delta^N(x;x')\Big|_{x=x'}=&B_N(k_0^2\eta^2)^{2\nu}\frac{(-1)^N 4^\nu}{\sqrt{\pi}}\frac{\Gamma^2(1+\nu)\Gamma(N-\frac 1 2+\nu)}{\Gamma(N)\Gamma(N+\nu)\Gamma(N+2\nu)}\eta^{2(N-1)}\\
    &\times\Bigg(-\frac{1}{D-1}\bar{\eta}_{\rho\sigma}+\frac{N-1+2\nu}{N}\delta_\rho^0\delta_\sigma^0\Bigg)
\,,
    \end{split}
\end{equation}
resulting in the following contribution to the stress-energy tensor,
\begin{eqnarray}
\langle\Omega|T_{\mu\nu}|\Omega\rangle^N
  &=& -\frac{H^D|1-\epsilon|^D}{2(4\pi)^{D/2}}
        \,\frac{(z_0^2)^{N+\frac{D-1}{2}+\nu}}{N+\frac{D-1}{2}+\nu}
\,\frac{\Gamma(1\!-\!N\!-\!\nu)\Gamma(1\!-\!N\!-\!2\nu)}
     {\Gamma(\frac{3}{2}\!-\!N\!-\!\nu)\Gamma(\frac{D+1}{2})\Gamma(N\!+\!1)}
\label{Tmn:N>}
\\
&\times&\!\!
\bigg[\Big((D\!-\!1)(1\!-\!2\nu)\!-\!(D\!-\!2)N\Big)a^2\delta_\mu^0\delta_\nu^0
        + \frac12
        \Big((D\!-\!1)(1\!-\!2\nu)\!-\!2(D\!-\!2)\!N\Big)g_{\mu\nu}
   \bigg]\,.
\nonumber
\end{eqnarray}
%

\subsection{Renormalization}

 The total one-loop stress energy tensor is the sum of the
three contributions~(\ref{Tmn:infty}), (\ref{Tmn:N<}) and~(\ref{Tmn:N>}),
\begin{equation}
  \langle\Omega|T_{\mu\nu}|\Omega\rangle
    =   \langle\Omega|T_{\mu\nu}|\Omega\rangle_{\infty}
    +   \sum_{N=0}^\infty\langle\Omega|T_{\mu\nu}|\Omega\rangle_N
    +   \sum_{N=0}^\infty\langle\Omega|T_{\mu\nu}|\Omega\rangle^N
\,.
\label{Tmn:total}
\end{equation}
Note that the ultraviolet divergence (which in dimensional regularization
appears as a term multiplying $1/(D-4)$) is confined
to~(\ref{Tmn:infty}). Indeed, when expanded around $D=4$,
Eq.~(\ref{Tmn:infty}) gives,
\begin{eqnarray}
\langle\Omega|T_{\mu\nu}|\Omega\rangle_{\infty}
  &=& \bigg\{-\frac{(2-\epsilon)\epsilon H^D}{8\pi^2}\,\frac{1}{D\!-\!4}
\nonumber\\
  &&-\frac{(2\!-\!\epsilon)\epsilon H^4}{16\pi^2}
   \bigg[4-\epsilon+\gamma
    +\ln\bigg(\frac{(1-\epsilon)^2}{4\pi}\bigg)
        +\psi\Big(\frac{1}{1-\epsilon}\Big)
        +\psi\Big(-\frac{\epsilon}{1-\epsilon}\Big)
   \bigg]
     \bigg\}
\nonumber\\
&&
\times\bigg[\epsilon a^2\delta_\mu^0\delta_\nu^0
      +\Big(\epsilon-\frac{3}{4}\Big)g_{\mu\nu}
\bigg]
 +\frac{(2-\epsilon)\epsilon H^4}{128\pi^2}g_{\mu\nu}
 + {\cal O}(D\!-\!4)
\,.
\label{Tmn:infty2}
\end{eqnarray}
 It is known that this theory can be renormalized by
the $R^2$ counterterm only. Indeed, taking a functional derivative
with respect to $g^{\mu\nu}$ of the counterterm action results
 in
\begin{equation}
  -\frac{2}{\sqrt{-g}}\frac{\delta}{\delta g^{\mu\nu}}\int d^D x\sqrt{-g}
         \alpha R^2
= \alpha (4\nabla_\mu\nabla_\nu R - 4g_{\mu\nu}\square R
          + g_{\mu\nu} R^2 -4RR_{\mu\nu})
\,.
\end{equation}
Making use of the corresponding expressions for $R$ and $R_{\mu\nu}$ in
FLRW spaces (see {\it e.g.} Ref.~\cite{JMP}) this evaluates to
\begin{eqnarray}
 -\frac{2}{\sqrt{-g}}\frac{\delta}{\delta g^{\mu\nu}} \int d^D x\sqrt{-g} R^2
  &=&144(2-\epsilon)\epsilon
   H^4\bigg[\epsilon\, a^2\delta_\mu^0\delta_\nu^0
     +\Big(\epsilon-\frac{3}{4}\Big)g_{\mu\nu}\bigg]
\nonumber\\
 &&\hskip -3.2cm
-H^4\Big(48\epsilon(1-4\epsilon+\epsilon^2)a^2\delta_\mu^0\delta_\nu^0-12(3-22\epsilon+22\epsilon^2-4\epsilon^3)g_{\mu\nu}\Big)(D\!-\!4)\qquad\quad
\nonumber\\
    &&\hskip -3.2cm
  +\,\mathcal{O}\Big((D\!-\!4)^2\Big)\,.
\label{R2:varied}
\end{eqnarray}
Note that the $H^D$ term in Eq.~(\ref{Tmn:infty2}) can be expanded as,
\begin{equation}
 H^D =  H^4 \mu^{D-4} \left[1+\frac{D\!-\!4}{2}\ln\Big(\frac{H^2}{\mu^2}\Big)
                      \right]
     + {\cal O}((D\!-\!4)^2)
\,,
\end{equation}
where $\mu$ is an arbitrary renormalization scale. From
Eqs.~(\ref{Tmn:infty2}) and~(\ref{R2:varied})
we see that the divergence in~(\ref{Tmn:infty2}) is canceled by
\begin{equation}\label{alpha}
    \alpha=\frac{\mu^{D-4}}{1152\pi^2(D\!-\!4)}
\,,
\end{equation}
where $\mu$ controls the undetermined finite part of $\alpha$.
The renormalized stress-energy tensor can be now easily obtained
\begin{equation}\label{Tmnfinal}
    \begin{split}
        \langle\Omega|T_{\mu\nu}|\Omega\rangle&=-\frac{H^4}{48\pi^2}
  \bigg\{(2+16\epsilon-16\epsilon^2+3\epsilon^3)a^2\delta_\mu^0\delta_\nu^0
 -\frac{1}{8}(12+62\epsilon-215\epsilon^2+146\epsilon^3-24\epsilon^4)
        g_{\mu\nu}\\
        &+3\epsilon(2-\epsilon)\bigg(\gamma+\psi\Big(-\frac{\epsilon}{1-\epsilon}\Big)+\psi\Big(\frac{1}{1-\epsilon}\Big)+\ln\Big(\frac{(1-\epsilon)^2
        H^2}{4\pi\mu^2}\Big) \bigg)
\\&\qquad\qquad\times
 \bigg[\epsilon\, a^2\delta_\mu^0\delta_\nu^0+\Big(\epsilon-\frac{3}{4}\Big)g_{\mu\nu}\bigg]\bigg\}\\
        &+\sum_{N=0}^\infty \left(\langle\Omega|T_{\mu\nu}|\Omega\rangle_{N}
            +\langle\Omega|T_{\mu\nu}|\Omega\rangle^{N}\right)_{D\rightarrow 4}
\,,
   \end{split}
\end{equation}
where the terms in the last line denote the $D\rightarrow 4$ limit
of Eqs.~(\ref{Tmn:N<}) and~(\ref{Tmn:N>}).


The sums over $N$ can be performed and the
final result can than be recast in terms of a hypergeometric
function ${}_2 F_3$. This is a useful procedure for studying the ultraviolet
behavior of the corrections (when the cutoff $|z_0|\rightarrow \infty$).
Since here we are primarily interested in the infrared sector
$|z_0|\ll 1$, we shall not perform the summation over $N$.

\subsection{Resolving the divergencies of the digamma functions}

 Even though the ultraviolet divergences have been
removed by dimensional renormalization, the
renormalized stress-energy~(\ref{Tmnfinal})
still seems to diverge at the poles of
the (di)gamma functions~(\ref{lower}--\ref{upper})
(see also Eqs.~(\ref{poles:lowerN}--\ref{poles:upperN})).
We shall now show that these divergences
are only apparent however, and that they are canceled by
the correction terms in~(\ref{Tmnfinal}) given by~(\ref{Tmn:N<})
and~(\ref{Tmn:N>}), precisely as there were designed to do.
\footnote{One might think that the second
digamma function in~(\ref{Tmnfinal}) has a simple pole also at $\epsilon=2$,
but that is canceled by the $(2-\epsilon)$ prefactor.}

Let us first consider the case $\epsilon<1$. Expanding~(\ref{lower})
around the $N$-th pole we have
\begin{equation}
    -\frac{\epsilon}{1-\epsilon}=-N +\delta
\quad \Rightarrow\quad
   \epsilon = \frac{N-\delta}{N+1-\delta}
\,,\qquad
    \nu=\frac32+N-\delta
\,,
\label{delta:lower}
\end{equation}
where $N$ is a positive integer and $\delta$ is an infinitesimal quantity.
In this case the first digamma function in~(\ref{Tmnfinal})
diverges, and its contribution to the stress-energy tensor is
\begin{equation}
    \frac{H^4}{16\pi^2}\frac{N(N+2)}{(N+1)^2}\frac{1}{\delta}
        \Big[\epsilon a^2\delta_\mu^0\delta_\nu^0
                +\Big(\epsilon-\frac{3}{4}\Big)g_{\mu\nu}\Big]
        + {\cal O}(\delta^0)
\,,
\label{TmnInf:pole:<}
\end{equation}
where we used
\begin{equation}
    \psi(-N+\delta)=-\frac{1}{\delta}+\mathcal{O}\big(\delta^0\big).
\end{equation}
To check that our construction works next we rewrite the $N$-th
term~(\ref{Tmn:N<}) by using~(\ref{delta:lower})
\begin{eqnarray}
 \langle\Omega|T_{\mu\nu}|\Omega\rangle_N
      &\stackrel{D\rightarrow 4}{\longrightarrow}&
    -\frac{H^4}{32\pi^2(1\!+\!N\!-\!\delta)^4}\frac{(z_0^2)^\delta}{\delta}
      \frac{\Gamma(\frac{5}{2}\!-\!\delta)\Gamma(N\!+\!3\!-2\delta)}
           {\Gamma(3\!-\!\delta)\Gamma(\frac{5}{2})\Gamma(N)}
\nonumber\\
 &&\times\, \Big[2(2N\!-\!\delta) a^2\delta_\mu^0\delta_\nu^0
              +(N\!-\!3\!+\!\delta)g_{\mu\nu}\Big]
\,.
\label{Tmn:N<:pole}
\end{eqnarray}
When expanded in powers of $\delta$ this gives,
\begin{eqnarray}
 \langle\Omega|T_{\mu\nu}|\Omega\rangle_N
     & \stackrel{D\rightarrow 4}{\longrightarrow}&
    -\frac{H^4}{16\pi^2}\frac{N(N\!+\!2)}{(N\!+\!1)^2}
\left[\frac{1}{\delta}+\ln(z_0^2)
           -\psi\Big(\frac52\Big)
           -2\psi\big(N+3\big)
           +\psi\big(3\big)
           +\frac{4}{N\!+\!1}
                      \right]
\hskip -0.8cm
\nonumber\\
 &&\hskip .cm \times
\, \bigg[\frac{N}{N\!+\!1}\, a^2\delta_\mu^0\delta_\nu^0
        +\frac{N\!-\!3}{4(N\!+\!1)}\,g_{\mu\nu}
       \!-\!\frac{\delta}{2(N\!+\!1)}
              \Big(a^2\delta_\mu^0\delta_\nu^0\!-\!\frac12g_{\mu\nu}\Big)
  \bigg]
+ {\cal O}(\delta)
\,.\qquad
\label{Tmn:N<:pole2}
\end{eqnarray}
Note that close to the pole $N/(N\!+\!1)=\epsilon+{\cal O}(\delta)$
and $(N\!-\!3)/[4(N\!+\!1)]=\epsilon\!-\!(3/4)+{\cal O}(\delta)$,
such that the
tensor structures in~(\ref{Tmn:N<:pole2}) and~(\ref{TmnInf:pole:<})
are (to leading order in $\delta$) identical.
Moreover, by comparing Eq.~(\ref{Tmn:N<:pole2}) with~(\ref{TmnInf:pole:<})
we see that the
${\cal O}(1/\delta)$ terms cancel, as required.
The resulting leading order contribution
to the one-loop stress-energy tensor is finite and
depends logarithmically on the scale factor as
$\propto H^4a^2[\,\ln(a)\!+\!{\rm const.}]$, where
the constant term contains a logarithm of the cutoff $k_0$.

%

\bigskip

In the case when $\epsilon>1$ the singularities of the digamma
functions in the stress-energy tensor~(\ref{Tmnfinal})
conform with the poles~(\ref{upper}). As above, expanding around the poles
\begin{equation}
    \frac{1}{1-\epsilon}=-2-N+\delta
\quad \Rightarrow\quad \nu = -\frac32-N+\delta
\,,\qquad \epsilon = \frac{N+3-\delta}{N+2-\delta}
\,,
\end{equation}
we obtain the following contribution to~(\ref{Tmnfinal})
from the digamma function:
\begin{equation}
 \frac{H^4}{16\pi^2}\frac{(N+1)(N+3)}{(N+2)^2}\frac{1}{\delta}
    \Big(\epsilon\, a^2\delta_\mu^0\delta_\nu^0
           +\Big(\epsilon-\frac{3}{4}\Big)g_{\mu\nu}\Big)
 +{\cal O}(\delta^0)
\,.
\label{TmnInf:pole:>}
\end{equation}
On the other hand, the correction~(\ref{Tmn:N>})
contributes the following to the scalar stress-energy~(\ref{Tmnfinal})
\begin{eqnarray}
 \langle\Omega|T_{\mu\nu}|\Omega\rangle^N
     & \!\!\! \stackrel{D\rightarrow 4}{\longrightarrow} \!\!\! &
    -\frac{H^4}{16\pi^2}\frac{(N\!+\!1)(N\!+\!3)}{(N\!+\!2)^2}
\left[\frac{1}{\delta}+\ln(z_0^2)
           -\psi\Big(\frac52\Big)
           -2\psi\big(N\!+\!4\big)
           +\psi\big(3\big)
           +\frac{4}{N\!+\!2}
                      \right]
\nonumber
\\
 &&\!\!\!\!\times \bigg[\frac{N\!+\!3}{N\!+\!2}\, a^2\delta_\mu^0\delta_\nu^0
        +\frac{N\!+\!6}{4(N\!+\!2)}\,g_{\mu\nu}
       -\frac{3\delta}{2(N\!+\!2)}
              \Big(a^2\delta_\mu^0\delta_\nu^0+\frac12g_{\mu\nu}\Big)
  \bigg] + {\cal O}(\delta)
\,.\qquad
\label{Tmn:N>:pole2}
\end{eqnarray}
Noting that here $(N\!+\!3)/(N\!+\!2)=\epsilon+{\cal O}(\delta)$
and $(N\!+\!6)/[4(N\!+\!2)]=\epsilon\!-\!(3/4)+{\cal O}(\delta)$, we see
that the pole contributions ${\cal O}(1/\delta)$
in~(\ref{Tmn:N>:pole2}) and~(\ref{TmnInf:pole:>})
cancel, resulting again in a finite contribution to the scalar
one-loop stress-energy of the form,
$\propto H^4a^2[\,\ln(a)\!+\!{\rm const.}]$.

\section{Discussion}

We studied the propagator of the massless, minimally coupled
scalar on a spatially flat, FLRW background of arbitrary dimension
and constant $\epsilon = -\dot{H}/H^2$. A previous result
(\ref{problem}), first derived \cite{JMP,JP2} by solving the
propagator equation (\ref{propeqn}), was seen to agree with the
Bunch-Davies mode sum for infinite space. It diverges at discrete
values of $\epsilon$ (\ref{lower}-\ref{upper}), even away from
coincidence and with the dimensional regularization in effect. The
origin of these divergences derives from the way dimensional
regularization treats the infrared divergences that the mode sum
possesses for all $\epsilon$ in the range $0 \leq \epsilon \leq
2(D\!-\!1)/D$. For most values of $\epsilon$ in this range the
infrared divergences are of the power law type that are set to
zero by the automatic subtraction of dimensional regularization.
The infrared divergences become logarithmic for the special values
of $\epsilon$ given by the two series (\ref{lower}) and
(\ref{upper}), which is why the infinite space propagator
(\ref{problem}) diverges for these values.

Of course one should never subtract infrared divergences, even power
law ones. Infrared divergences signify an unphysical feature of
whatever question is being posed and the correct way to avoid them
is by making appropriate changes in the question. The unphysical
feature of the infinite space propagator is that no local observer
could prepare the initial state in coherent Bunch-Davies vacuum over
more than a Hubble volume. A more realistic situation is attained
either by preparing the initially super-horizon modes in some less
singular state or else by simply not having any initially
super-horizon modes.

We implemented the latter fix, making the integral approximation
to the discrete mode sum but with a nonzero lower limit. The
resulting propagator (\ref{newsum}) could be expressed as the old
result (\ref{problem}), minus a series of homogeneous solutions.
Canceling the infrared divergences requires only the most
singular contributions. For the inflationary case of $0 \leq
\epsilon < 1$ our results for the corrections are (\ref{ansatz})
and (\ref{akl}); for the decelerating case of $1 < \epsilon \leq
2(D\!-\!1)/D$ they are (\ref{bansatz}) and (\ref{bkl}). We showed
explicitly that the first two divergences of (\ref{lower}) and the
first two divergences of (\ref{upper}) are canceled in this way,
and we obtained the resulting propagators with dimensional
regularization still in effect. We also showed that these results
agree with special cases which have been reported in the
literature \cite{OW,ITTW}. Moreover as an example we calculated
the one-loop expectation value of the stress-energy tensor and
showed that the divergences as discussed in \cite{JP2} are
canceled and replaced with a term that grows as the logarithm of
the infrared cutoff.

The case of $\epsilon = 1$ deserves special comment. From the
convergence of the lower singularities (\ref{lower}) and the
upper ones (\ref{upper}) it might seem that the infinite space
mode sum is highly infrared divergent. However, this is just an
illusion derived from the fact that our convention for the zero
of conformal time shifts discontinuously at $\epsilon = 1$. If
one takes the limit $\epsilon \rightarrow 1$ while holding fixed
the co-moving time and the initial Hubble parameter then the mode
functions have a perfectly well-behaved form,
\begin{equation}
\lim_{\epsilon \rightarrow 1} u(t,k) = \frac{a^{1-\frac{D}2}}{\sqrt{2 H_0}}
\Biggl[\frac{k^2}{H_0^2} \!-\! \Bigl(\frac{D \!-\!2}2\Bigr)^2\Biggr]^{-\frac14}
\!\! \exp\Biggl[-i \sqrt{\frac{k^2}{H_0^2} \!-\! \Bigl(\frac{D \!-\! 2}2
\Bigr)^2 } \, \ln(a)\Biggr] .
\end{equation}
The propagator also has a smooth limit,
\begin{equation}
\lim_{\epsilon \rightarrow 1} i\Delta_{\infty}(x;x') =
\frac{[H H']^{\frac{D}2-1}}{(4\pi)^{\frac{D}2}} \,
\Gamma\Bigl(\frac{D}2 \!-\!1\Bigr) \Bigl(\frac{4}{Y}\Bigr)^{\frac{D}2 -1} \; ,
\end{equation}
where we define the symbol $Y$ as
\begin{equation}
Y(x;x') \equiv H_0^2 \Delta x^2 - \Biggl(\Bigl\vert \ln\Bigl(\frac{a}{a'}
\Bigr) \Bigr\vert - i \varepsilon \Biggr)^2 \; .
\end{equation}

An important distinction exists between the inflationary case of
$0\leq\epsilon<1$ and the case of deceleration. The reason for
this is that the parameter used to characterize the infrared
cutoff, $z_0$, approaches zero in an inflationary space, but it
grows without bound in a decelerating space-time. In an
inflationary space-time the physical wavelength associated with
the infrared cutoff, $a(t)/k_0$, grows faster than the Hubble
radius. Thus a super-Hubble cutoff stays super-Hubble at all
times. In contrast, for a decelerating universe the Hubble radius
grows faster than $a(t)/k_0$ and therefore an initial super-Hubble
cutoff will eventually enter the Hubble radius. Hence the effect
of the cutoff becomes more and more profound as time evolves. This
behavior can be seen nicely from the expectation value of the
stress-energy tensor (\ref{Tmnfinal}). The corrections from
$\delta i\Delta_N$ and $\delta i\Delta^N$ have the following
respective time dependences:
\begin{equation}
    \begin{split}
     H^4(z_0^2)^{N-\frac{\epsilon}{1-\epsilon}}&\propto
     a^{-2[\epsilon+N(1-\epsilon)]}\\
     H^4(z_0^2)^{N+\frac{3-2\epsilon}{1-\epsilon}}&\propto
     a^{-2[3+N(1-\epsilon)]}\,.
     \end{split}
\label{}
\end{equation}
This behavior should be compared to the tree level contribution to
the equation of motion, which follows from the Friedmann equations,
and scales as $H^2\propto a^{-2\epsilon}$ and $\dot H\propto a^{-2\epsilon}$.
Thus we see that for an
inflationary space, $\epsilon<1$,
all corrections $N\geq 1$ decay faster with time
than the tree-level contributions, while
the $N=0$ correction scales with time equally as $H^2$ and  $\dot H$.
For a decelerating space $\epsilon>1$ however, there
are infinitely many values of $N$ for which both corrections grow
indefinitely as the universe
expands.~\footnote{In order to get the correct late time
behavior in this case,
one would have to sum the series in $N$ and consider
the asymptotic behavior of the resulting
hypergeometric function ${}_2F_3$ in the limit when $|z_0|\gg 1$.
Since that limit corresponds to a sub-Hubble cutoff,
it is not of great interest to us.}


Thus we conclude that for an inflationary universe, the initial
conditions become less visible as time progresses. In
this case it does not matter much which of the two fixes for the
infrared problems of Section 3 is employed, or precisely how it is
implemented, while the opposite is true for a decelerating
universe. Although we have worked out the fix based on a finite
spatial manifold, the absence of evidence for a finite size to the
universe suggests that the more physically relevant fix for
deceleration is the one based on a less singular ensemble of
initially super-horizon modes \cite{AV}. Because the initial
condition becomes progressively more visible for deceleration, the
physically relevant choice is probably the one consistent with at
least 60 e-foldings of primordial inflation. Note that modes which
experience first horizon crossing will be {\it even more} infrared
singular than for Bunch-Davies vacuum. It would only be for
currently unobservable super-horizon modes that one could use the
less singular mode functions needed to regulate the infrared
divergence. Hence one expects the infrared behavior of the
physically relevant propagator for deceleration to be strongly
influenced by the initial condition provided by primordial
inflation.

\vskip 1cm

 \centerline{\bf Acknowledgements}

We are grateful for conversations on this subject with N. C. Tsamis.
This work was partially supported by FOM grant 07PR2522, by Utrecht
University, by NSF grant PHY-0653085, and by the Institute for
Fundamental Theory at the University of Florida.

\end{document}